\documentclass[preprint,12pt]{elsarticle}

\usepackage{amsmath, amsthm, amsfonts, amssymb}
\usepackage{graphicx}
\biboptions{sort&compress}

\numberwithin{equation}{section}

\newcommand\Poincare 	{Poincar\'e\ }
\newcommand\Schrodinger {Schr\"{o}dinger }
\newcommand\Tr 			{\mathrm{Tr}}

\journal{Nuclear Physics B}

\begin{document}

\begin{frontmatter}



\title{Entanglement Entropy and Duality in AdS$_4$}


\author[ntua]{Ioannis Bakas}
\ead{bakas@mail.ntua.gr}
\author[ntua]{Georgios Pastras\corref{cor}}
\ead{pastras@mail.ntua.gr}

\address[ntua]{Department of Physics, School of Applied Mathematics and Physical Sciences\\National Technical University, Athens 15780, Greece}

\cortext[cor]{Corresponding author}

%This article is registered under preprint number: 1503.00627

\begin{abstract}

Small variations of the entanglement entropy $\delta S$ and the expectation value of the modular Hamiltonian $\delta E$
are computed holographically for circular entangling curves in the boundary of $AdS_4$, using gravitational perturbations with general
boundary conditions in spherical coordinates. Agreement with the first law of thermodynamics, $\delta S = \delta E$, requires that the
line element of the entangling curve remains constant. In this
context, we also find a manifestation of electric-magnetic duality for the entanglement entropy and the corresponding modular Hamiltonian,
following from the holographic energy-momentum/Cotton tensor duality.

\end{abstract}


\end{frontmatter}



\section{Introduction}
\label{sec:introduction}
It has been known for a long time that there is a striking similarity between black hole physics and thermodynamics \cite{Bekenstein:1973ur,Bardeen:1973gs,Hawking:1974sw}, which suggests a deep connection between gravity and thermodynamics.
This led to several attempts to understand Einstein's equations as effective equations emerging from the thermodynamics of
underlying degrees of freedom, such as \cite{Jacobson:1995ab}.
On the other hand, AdS/CFT correspondence \cite{Maldacena:1997re,Gubser:1998bc,Witten:1998qj} provides a broad framework
allowing the description of gravitational theories with AdS asymptotics in $d+1$ dimensions as emergent from strongly coupled
conformal field theories in $d$ dimensions. A fair question in the AdS/CFT framework is whether the similarities between
gravity and thermodynamics can be explained by considering Einstein's equations as thermodynamic relations for the
conformal field theory degrees of freedom \cite{Verlinde:2010hp}.

More recently, it has also been suggested that the connection between gravity and thermodynamics should not be attributed
to thermal statistics, but rather to quantum statistics related to quantum entanglement
physics \cite{Ryu:2006bv,Ryu:2006ef,Hubeny:2007xt,Nishioka:2009un,VanRaamsdonk:2009ar,VanRaamsdonk:2010pw,Takayanagi:2012kg}.
More specifically, it has been conjectured that the entanglement entropy, which is a measure of entanglement between subsystems of
a composite quantum system and it is defined for a given entangling surface that separates the degrees of freedom of the
corresponding conformal field theory into two subsystems, is directly connected to the area of an open extremal hypersurface
in the emergent asymptotically AdS geometry whose boundary is the entangling surface.
This conjecture, which is named after Ryu-Takayanagi \cite{Ryu:2006bv,Ryu:2006ef}, provides a quantitative tool to understand
how gravitational dynamics
emerges from thermodynamics related to entanglement in the boundary conformal field theory.

So far, this programme has been advanced by comparing the variation of entanglement entropy to the variation of the expectation
value of the so called modular Hamiltonian for any given entangling surface \cite{Blanco:2013joa,Wong:2013gua}. The latter can
be expressed in terms of the holographic energy-momentum tensor when spherical entangling surfaces are taken in \Poincare
coordinates \cite{Casini:2011kv}, while the former is provided by the Ryu-Takayanagi formula \cite{Ryu:2006bv,Ryu:2006ef}.
Enforcing the first law of thermodynamics for entanglement through holography, imposes constraints for the metric perturbations around AdS space,
which to linear order turn out to be Einstein's equations satisfying Dirichlet boundary conditions \cite{Lashkari:2013koa,Faulkner:2013ica}.
In this context, it is also known that all solutions of the linearized Einstein's equations with Dirichlet boundary conditions
satisfy the first law of thermodynamics.

In this paper, we work out the holographic realization of the first law of thermodynamics for small perturbations of spherical $AdS_4$
space-time having axial symmetry and satisfying general boundary conditions. The general framework for field equations satisfying
general boundary conditions in $AdS_4$ is provided in reference \cite{ishi}. The entangling surfaces are now curves, since the boundary
conformal field theory is $2+1$ dimensional, and they are taken to be circular, bounding a polar cap region, so that they respect the
axial symmetry of the bulk geometry. Our general result is that the first law of thermodynamics for entanglement is realized
holographically in all cases, hereby extending previous works beyond Dirichlet boundary conditions, provided that the line element of
the entangling curve is inert to the perturbations.

In this context, we also examine the role of gravitational electric-magnetic duality to entanglement physics. It is well known that
small perturbations of maximally symmetric spaces exhibit a rank-2 generalization of electric-magnetic duality, interchanging the
linearized Einstein equations with the Bianchi identities in four space-time dimensions. This symmetry was originally discussed
for gravitons in Minkowski space \cite{Henneaux:2004jw}, but it was subsequently generalized in the presence of cosmological
constant by considering metric perturbations of $dS_4$ \cite{julia} and $AdS_4$ space-time
\cite{Leigh:2007wf,deHaro:2008gp,Bakas:2008zg,Bakas:2009da}. There, it was also found that electric-magnetic duality in $AdS_4$ has
a holographic manifestation as energy-momentum/Cotton tensor duality. These considerations provide the gravitational analogue
of the holographic interpretation of electric-magnetic duality in theories with $U(1)$ gauge symmetry \cite{Witten:2003ya},
but their physics in the space of boundary three-dimensional conformal theories still remain largely unexplored.

Duality acts on metric perturbations by interchanging their boundary conditions. Hence, our interest in the holographic description
of gravitational perturbations satisfying general boundary conditions in the spirit of reference \cite{Compere:2008us}.
Extending the applications of gravitational duality to entanglement entropy and related issues may shed new light into this
interesting subject.

The material of this paper is organized as follows: In section 2, we present a review of the notions of entanglement
entropy and modular Hamiltonian, together with their holographic description, and include various formulae that will be used in the
computations. In section 3, we discuss the general theory of gravitational perturbations of $AdS_4$ space-time and
formulate the linearized Einstein equations as an effective \Schrodinger problem, splitting the perturbations into
two distinct classes with opposite parity. The presentation is made general, encompassing arbitrary boundary conditions.
In section 4, we compute holographically the variations of the entanglement entropy and the modular Hamiltonian and compare
the two expressions for general boundary conditions. The first law of thermodynamics for entanglement is verified in all cases, while describing the
subtleties that go into the computation. In section 5, we address the role of electric-magnetic duality in holography and
study its implications for the first law of thermodynamics for entanglement. Finally, section 6, contains our conclusions
and a short discussion of open problems. There are also three appendices containing various technical details and formulae
that are used in the main text.

\section{Entanglement Entropy and Holography}
\label{sec:entanglement_holography}
We present a brief account of the notions of entanglement entropy and modular Hamiltonian together with their holographic description
in terms of bulk space geometry. We also derive some general formulae that will be used later for gravitational perturbations
of $AdS_4$ space-time satisfying general boundary conditions.

\subsection{First Law of Thermodynamics for Entanglement}
\label{subsec:EE_and_nodular_Hamiltonian}
Consider a composite quantum system comprising of several subsystems. Even if the composite system lies in a pure state, with density matrix $\rho$,
this may not be true for its subsystems, which are hereby described by a density matrix equal to the partial trace of $\rho$ over the
degrees of freedom of the complementary subsystem,
\begin{equation}
\rho_A = \Tr_{A^C} \rho.
\label{eq:EEMH_density_matrix}
\end{equation}
When the complementary subsystems $A$ and $A^C$ are not entangled, the reduced density matrix $\rho_A$ also describes a
pure state. Entanglement between systems $A$ and $A^C$ is encoded to the spectrum of the reduced density matrix $\rho_A$
under the implicit assumption that the composite system lies in a pure state. The entanglement entropy is defined
as the von Neumann entropy associated to the reduced density matrix $\rho_A$,
\begin{equation}
{S_A} := - \Tr \left( \rho_A \ln \rho_A \right) .
\label{eq:EEMH_entanglement_entropy}
\end{equation}

The density matrix $\rho_A$ is Hermitian and positive semi-definite, leading to the definition of the corresponding
modular Hamiltonian as
\begin{equation}
{\rho _A} := {e^{ - {H_A}}} .
\label{eq:EEMH_modular hamiltonian}
\end{equation}
Then, the entanglement entropy can be rewritten in terms of the modular Hamiltonian as
\begin{equation}
{S_A} =  - \Tr\left( {{\rho _A}\ln {\rho _A}} \right) = \Tr\left( {{\rho _A}{H_A}} \right) = \left\langle {{H_A}} \right\rangle .
\label{eq:EEMH_relation}
\end{equation}

Small variations in the pure state of the overall system or the region $A$ generate variations of the density matrix, $\delta {\rho _A}$,
and, thus, the entanglement entropy will also be perturbed. We have
\begin{equation}
\begin{split}
\delta {S_A} &=  - \Tr\left[ {\delta \left( {{\rho _A}\ln {\rho _A}} \right)} \right]\\
 & = - \Tr\left( {\ln {\rho _A}\delta {\rho _A}} \right) - \Tr\left( {\delta {\rho _A}} \right)\\
 & = \Tr\left( {{H_A}\delta {\rho _A}} \right) = \delta \left\langle {{H_A}} \right\rangle ,
\end{split}
\end{equation}
since the trace of the density matrix is normalized to one and thus, $\Tr\left( {\delta {\rho _A}} \right) = 0$.
Thus, the variations of the entanglement entropy and the expectation value of the corresponding modular Hamiltonian are equal
\begin{equation}
\delta {S_A} = \delta \left\langle {{H_A}} \right\rangle \equiv \delta E \,.
\label{eq:EEMH_first_law}
\end{equation}
This equation is the direct analog of the first law of thermodynamics for entanglement physics \cite{Blanco:2013joa,Wong:2013gua} that leads our work.

\subsection{Ryu-Takayanagi Conjecture}
\label{subsec:RT_conjecture}
The Ryu-Takayanagi conjecture \cite{Ryu:2006bv,Ryu:2006ef} connects the entanglement entropy of a region $A$ defined by the entangling
boundary surface $\partial A$ in the boundary field theory to the area of an extremal co-dimension two open surface in the bulk
gravitational dual theory with boundary $\partial A$. Specifically, the entanglement entropy is given by
\begin{equation}
{S_A} = {1 \over 4{G_N}} \, {\rm Area} \left(A^{{\rm{extr}}} \right) \, ,
\label{eq:RT_conjecture}
\end{equation}
where ${A^{{\rm{extr}}}}$ is the corresponding extremal co-dimension two surface in the bulk. In the following, without loss of generality,
we set Newton's gravitational constant $G_N = 1$.

These expressions are applicable to all holographic models. For $AdS_4$, which is of interest here, $\partial A$ is a closed
curve and ${A^{{\rm{extr}}}}$ is two-dimensional. Then, the area of the extremal surface, which itself will be denoted by ${A^{{\rm{extr}}}}$ in the
following, is expressed in terms of the induced
metric as
\begin{equation}
A^{\rm extr} = \int {{d^2}\sigma \sqrt \gamma } \, ,
\label{eq:extremal_surface_area}
\end{equation}
where
\begin{equation}
{\gamma_{ab}} = {g_{\mu \nu }}\frac{{\partial {X^\mu }\left( \sigma  \right)}}{{\partial {\sigma ^a}}}\frac{{\partial {X^\nu }\left( \sigma  \right)}}{{\partial {\sigma ^b}}}
\label{eq:induced_metric}
\end{equation}
and
\begin{equation}
\gamma = \det \left( {{\gamma_{ab}}} \right) .
\label{eq:induced_metric_determinant}
\end{equation}
Here, $g$ is the bulk metric, $\sigma^a$ are coordinates parametrizing the extremal surface,
$X\left( \sigma \right)$ are the parametric equations of the extremal surface in the bulk and $\gamma$ is the induced metric on the
extremal surface.

\subsection{Entanglement Entropy in Global AdS$_4$ for a Polar Cap Region}
\label{subsec:EE_pure_AdS4}
Extremal surfaces in AdS space-times have been mostly studied in \Poincare patch coordinates. In those coordinates, the space-time line
element of $AdS_4$ with unit scale is
\begin{equation}
d{s^2} = \frac{1}{{{z^2}}}\left( { - d{\tau ^2} + d{z^2} + d{x^2} + d{y^2}} \right) .
\label{eq:AdS_metric_Poincare}
\end{equation}
Then, for a disc region $A$ in the boundary plane described by ${\left(x - x_0 \right)^2} + {\left(y - y_0 \right)^2} \le {R^2}$,
the corresponding extremal surface ${A^{{\rm{extr}}}}$ in the AdS bulk is given by
\begin{equation}
\begin{split}
{\left(x - x_0 \right)^2} + {\left(y - y_0 \right)^2} + {z^2} &= {R^2} , \\
\tau &= \tau_0 \,.
\end{split}
\label{eq:extremal_surface_disc_poincare}
\end{equation}
Without loss of generality we may take the disc centered at $(0, \, 0)$.

Passing from \Poincare coordinates $\left(\tau,z,x,y\right)$ to global coordinates $\left(t,r,\theta,\phi\right)$ with
the aid of the coordinate transformation
\begin{equation}
\begin{split}
\tau  &= \frac{{\sqrt {{r^2} + 1} \sin t}}{{\sqrt {{r^2} + 1} \cos t + r\cos \theta }}\, , \\
z &= \frac{1}{{\sqrt {{r^2} + 1} \cos t + r\cos \theta }}\, , \\
x &= \frac{{r\sin \theta \cos \varphi }}{{\sqrt {{r^2} + 1} \cos t + r\cos \theta }}\, , \\
y &= \frac{{r\sin \theta \sin \varphi }}{{\sqrt {{r^2} + 1} \cos t + r\cos \theta }}\, ,
\end{split}
\label{eq:coordinate_transformation}
\end{equation}
the space-time metric takes the form
\begin{equation}
d{s^2} =  - \left( {{r^2} + 1} \right)d{t^2} + \frac{{d{r^2}}}{{{r^2} + 1}} + {r^2}\left( {d{\theta ^2} + {{\sin }^2}\theta d{\varphi ^2}} \right) ,
\end{equation}
whereas the extremal surface \eqref{eq:extremal_surface_disc_poincare} corresponding to the choice $(x_0, \, y_0) = (0, \, 0)$ is given in
global coordinates by
\begin{equation}
t = t_0, ~~~  r\left( \theta  \right) = \frac{1}{{\cos \theta \sqrt {{{\tan }^2}{\theta _0} - {{\tan }^2}\theta } }} ,\quad \theta  \in
\left[ {0,{\theta _0}} \right], \, \varphi  \in \left[ {0,2\pi } \right) \, ,
\label{eq:extremal_surface_global_r}
\end{equation}
where $\theta_0$ and $t_0$ are specific functions of $R$ and $\tau_0$ (see \ref{sec:coordinate_transformation} for more details).
Equivalently, we have parametrization
\begin{equation}
t = t_0, ~~~ \theta \left( r \right) = \arccos \left( {\cos {\theta _0}\sqrt {1 + \frac{1}{{{r^2}}}} } \right) ,\quad r  \in \left[ {\cot {\theta _0},\infty } \right), \, \varphi  \in \left[ {0,2\pi } \right).
\label{eq:extremal_surface_global_theta}
\end{equation}
The region $A$ becomes a polar cap in global coordinates and the complementary cap is the region $A^C$.

Figure 1 depicts the
regions $A$ and $A^C$ on the spherical boundary of space-time.
\newline
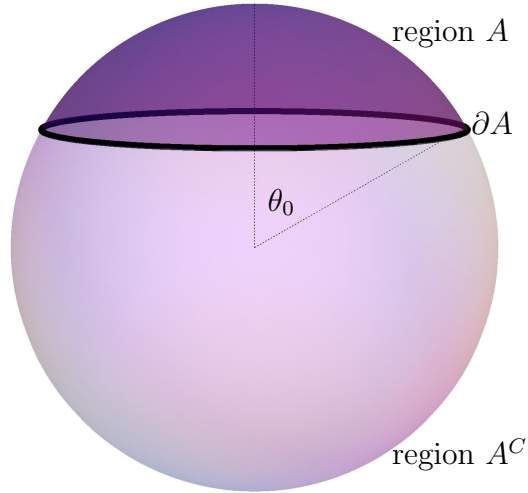
\begin{figure}[hb]
\begin{center}
\includegraphics[width=0.6\textwidth]{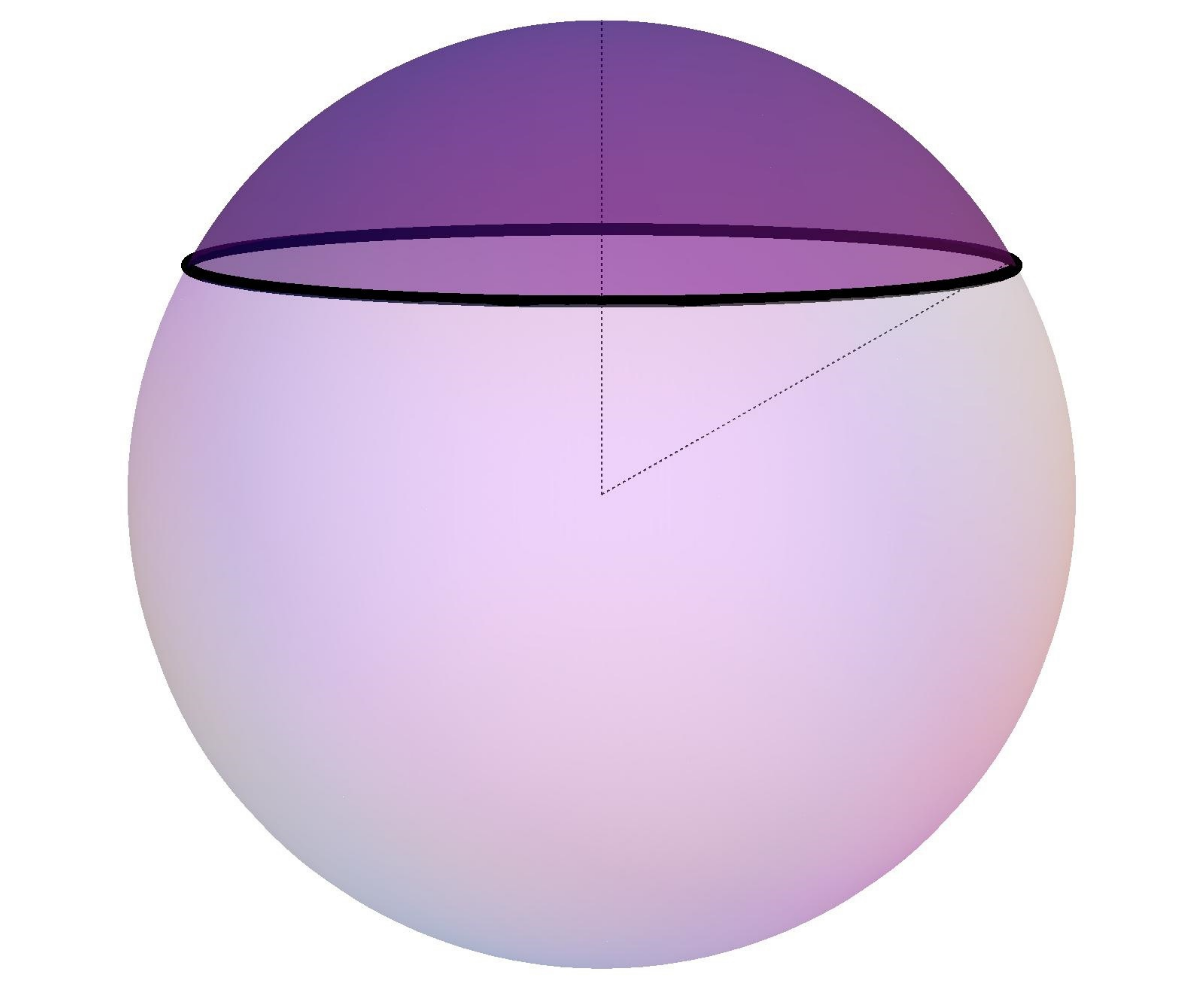}
\put(-112,111){$\theta_0$}
\put(-65,175){region $A$}
\put(-65,15){region $A^C$}
\put(-35,140){$\partial A$}
\end{center}
        \caption{The region $A$ has opening angle $\theta_0$. The entangling curve $\partial A$ is the thick line.}
        \label{fig:region}
\end{figure}

It is easy to confirm that the surface \eqref{eq:extremal_surface_global_theta} is extremal, obeying the particular conditions derived by
minimizing the area
functional \eqref{eq:extremal_surface_area}, under the assumption that the surface is rotationally symmetric, i.e.,
$t = t \left( r \right)$ and $\theta = \theta \left( r \right)$, since
\begin{align}
\left( \frac{{\left( {{r^2} + 1} \right)t'r\sin \theta }}{{\sqrt { - \left( {{r^2} + 1} \right)t{'^2} + \frac{1}{{{r^2} + 1}} + {r^2}\theta {'^2}} }} \right)' &= 0 \, ,\\
\left( \frac{{{r^3} \theta ' \sin \theta }}{{\sqrt { - \left( {{r^2} + 1} \right)t{'^2} + \frac{1}{{{r^2} + 1}} + {r^2}\theta {'^2}} }} \right)' &= r \cos \theta \sqrt { - \left( {{r^2} + 1} \right)t{'^2} + \frac{1}{{{r^2} + 1}} + {r^2}\theta {'^2}} \,  ,
\end{align}
where prime denotes differentiation with respect to $r$.

The extremal surface emanates from a polar cap boundary region described by $\theta \le \theta_0$. An introduction of non-vanishing
parameters $x_0$ or $y_0$ would rotate the entangling curve so that its symmetry axis would not anymore coincide with the axis corresponding
to the azimuthal angle $\phi$. Having said that, we stick to the choice $x_0 = y_0 = 0$ from now on.

Figure 2 depicts the extremal surface $A^{\rm{extr}}$ that emanates from a polar cap region in the boundary of space-time
in global coordinates and extends in the interior of space-time. The radial coordinate in the plot is proportional to the so called tortoise
coordinate, ${\rm arctan} r$.
\newline
\begin{figure}[hb]
\begin{center}
\includegraphics[width=0.6\textwidth]{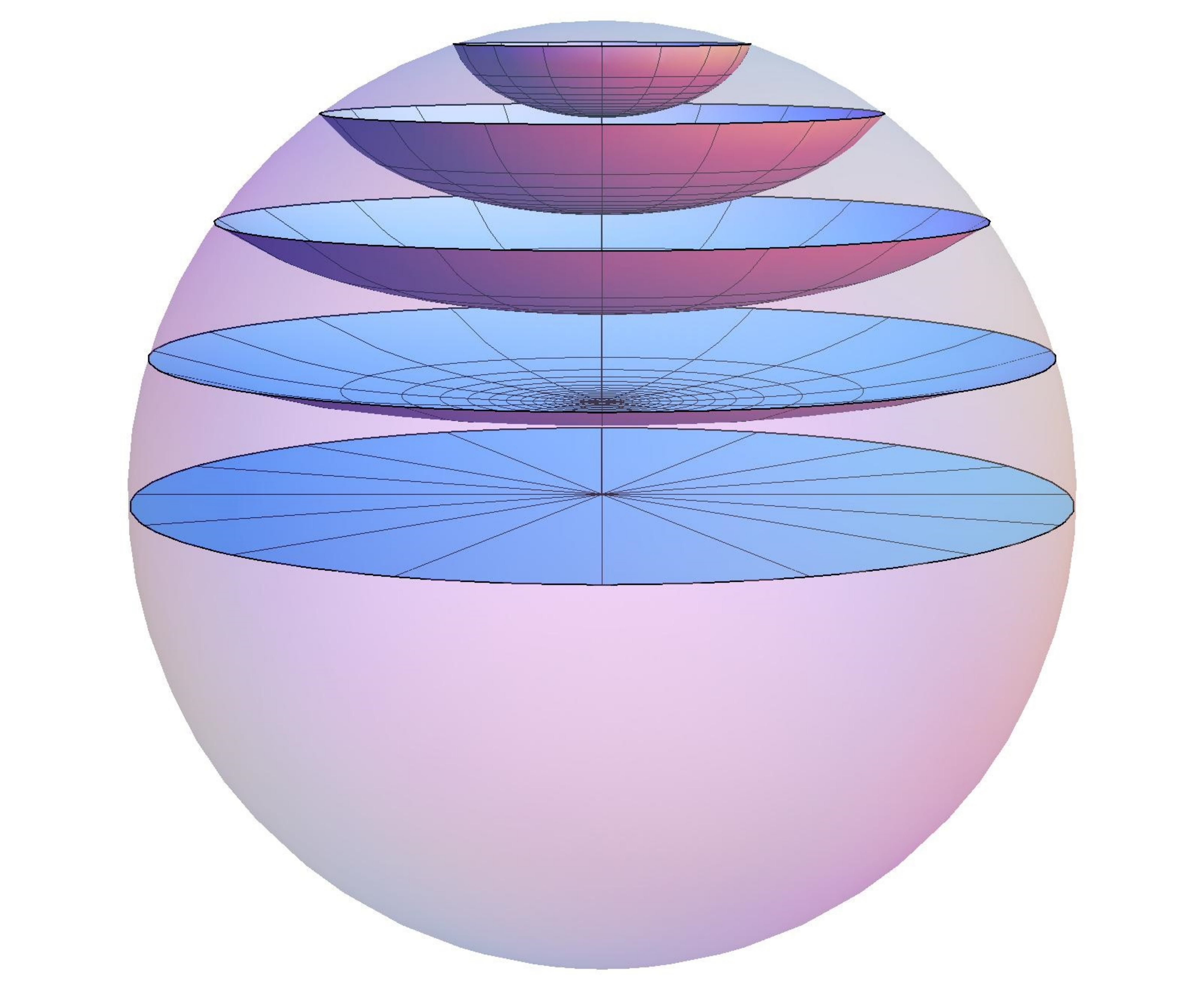}
\end{center}
        \caption{The extremal surfaces in $AdS_4$ space-time for various choices of $\theta_0$.}
        \label{fig:extremal}
\end{figure}

Parametrizing the extremal surface with the coordinates $\phi$ and $r$, the induced metric turns
out to be diagonal with elements
\begin{align}
{\gamma _{rr}} &= \frac{{{{\sin }^2}{\theta _0}}}{{{r^2}\left( {1 + \frac{1}{{{r^2}}}} \right)\left[ {1 - \left( {1 + \frac{1}{{{r^2}}}} \right){{\cos }^2}{\theta _0}} \right]}},\label{eq:unperturbed_induced_metric_rr}\\
{\gamma _{\varphi \varphi }} &= {r^2}\left[ {1 - \left( {1 + \frac{1}{{{r^2}}}} \right){{\cos }^2}{\theta _0}} \right].\label{eq:unperturbed_induced_metric_phiphi}
\end{align}
The determinant of the induced metric is $\gamma  = {\sin }^2{\theta _0}/(1 + 1/r^2)$ and the area follows from equation \eqref{eq:extremal_surface_area},
\begin{equation}
A^{\rm extr} = 2\pi \mathop {\lim }\limits_{r \to \infty } \left( {r\sin {\theta _0} - 1} \right) .
\label{eq:unperturbed_area}
\end{equation}
The first term is the divergent ``area law'' term, while the second one is universal independent of the UV cutoff\footnote{
Recall that the entanglement entropy for a disk region of radius $R$ in the boundary of $AdS_{d+1}$ is given by
\begin{equation}
{S_A} \sim
\begin{cases}
{a_1}{\left( {R/L} \right)^{d - 2}} + {a_3}{\left( {R/L} \right)^{d - 4}} +  \ldots  + {a_{d - 2}}\left( {R/L} \right) + {a_{d-1}}, & d\;\rm{odd},\\
{a_1}{\left( {R/L} \right)^{d - 2}} + {a_3}{\left( {R/L} \right)^{d - 4}} +  \ldots  + {a_{d - 3}}{\left( {R/L} \right)^2} + {a_{d-1}} {\rm log}
\left( {R/L} \right) , & d\;\rm{even},
\end{cases}
\label{eq:eeterms}
\end{equation}
where $L$ is the UV cutoff. The first term is the ``area law'' term. For $d$ even, the logarithmic term is universal and connected to the conformal anomaly \cite{Ryu:2006bv,Ryu:2006ef,Casini:2011kv,Myers:2010xs,Myers:2010tj,Solodukhin:2008dh}. For $d$ odd, which is the case of interest here ($d=3$), the constant term is universal and
it obeys a holographic ``c-theorem'' \cite{Myers:2010xs,Myers:2010tj}.
} \cite{Myers:2010xs,Myers:2010tj}.

\subsection{Perturbations of Entanglement Entropy}
\label{subsec:EE_perturbed_AdS4}

The area functional \eqref{eq:extremal_surface_area} depends on the background metric $g$ as well as on the embedding variables $X$, and, of course, it also depends implicitly on the entangling curve
$\partial A$ curving the region $A$ in the boundary field theory.
Perturbations of the bulk metric induce changes of the minimal surface. The variation of its area is given in general by
\begin{equation}
\delta A^{\rm extr} = {\left. {\frac{{\delta A^{\rm extr} \left( {g,X} \right)}}{{\delta g}}} \right|_{g = {g_0},\,X = {X_0}}}\delta g +
{\left. {\frac{{\delta A^{\rm extr} \left( {g,X} \right)}}{{\delta X}}} \right|_{g = {g_0},\,X = {X_0}}}\delta X
\end{equation}
as both the metric and the embedding equations of the extremal surface vary around their unperturbed values $g_0$ and $X_0$.

When the perturbation obeys Dirichlet boundary conditions, the region $A$ and its boundary $\partial A$ remain fixed. Then, the second
term vanishes, as the original surface described by equations $X_0$ extremizes the area functional with the given loop $\partial A$.
When the metric perturbations satisfy general boundary conditions, computing the variation of the area of the minimal surface is more subtle.
For this, we suppose that the boundary loop $\partial A$ specifying the extremal surface is determined by a set of parameters $b$,
which may also vary as metric perturbations are turned on. Then, the area functional has the particular form
\begin{equation}
A^{\rm extr}  = A^{\rm extr} \left( {g,X\left( {g,b} \right)} \right)
\label{eq:EE_functional}
\end{equation}
and its variation breaks down as follows,
\begin{multline}
\delta A^{\rm extr} = {\left. {\frac{{\delta A^{\rm extr} \left( {g,X} \right)}}{{\delta g}}} \right|_{g = {g_0},\,X = {X_0}}}\delta g \\
+ {\left. {\frac{{\delta A^{\rm extr} \left( {g,X} \right)}}{{\delta X}}} \right|_{g = {g_0},\,X = {X_0}}}
{\left. {\frac{{\delta X\left( {g,b} \right)}}{{\delta g}}} \right|_{g = {g_0},\,b = {b_0}}}\delta g \\
 + {\left. {\frac{{\delta A^{\rm extr} \left( {g,X} \right)}}{{\delta X}}} \right|_{g = {g_0},\,X = {X_0}}}
 {\left. {\frac{{\partial X\left( {g,b} \right)}}{{\partial b}}} \right|_{g = {g_0},\,b = {b_0}}}\delta b .
\end{multline}
The second term vanishes, as it corresponds to the variation of the area for variations of the extremal surface with
a fixed boundary that is provided by the entangling loop $\partial A$. The last term can be simplified if one considers the area
as functional of the metric as well as the parameters $b$ specifying the entangling loop. As a result, the total variation of the
area takes the form
\begin{equation}
\begin{split}
\delta A^{\rm extr} &= {\left. {\frac{{\delta A^{\rm extr} \left( {g,X} \right)}}{{\delta g}}} \right|_{g = {g_0},\,X = {X_0}}}\delta g +
{\left. {\frac{{\partial A^{\rm extr} \left( {g,b} \right)}}{{\partial b}}} \right|_{g = {g_0},\,b = {b_0}}}\delta b\\
 &:=
\delta {A_g^{\rm extr}} + \delta {A_b^{\rm extr}} \, ,
\end{split}
\end{equation}
hereby defining the individual contributions $\delta {A_g^{\rm extr}}$ and $\delta {A_b^{\rm extr}}$.

The first term follows by varying the relation \eqref{eq:extremal_surface_area} and the result is written in terms of the
unperturbed induced metric $\gamma_0$ and $\delta\gamma_{ab}$ as
\begin{equation}
\delta A_g^{\rm extr} = \frac{1}{2}\int {{d^2}\sigma \sqrt {{\gamma _0}} {\left( \gamma _0 \right)}^{ab}\delta {\gamma _{ab}}}  \, ,
\label{eq:dAg_formula}
\end{equation}
where
\begin{align}
{{\left( \gamma_0 \right)}_{ab}} = {\left( g_0 \right)}_{\mu \nu }\frac{{\partial {X_0}^\mu \left( \sigma  \right)}}{{\partial {\sigma ^a}}}\frac{{\partial {X_0}^\nu \left( \sigma  \right)}}{{\partial {\sigma ^b}}} , \\
\delta {\gamma _{ab}} = \delta {g_{\mu \nu }}\frac{{\partial {X_0}^\mu \left( \sigma  \right)}}{{\partial {\sigma ^a}}}\frac{{\partial {X_0}^\nu \left( \sigma  \right)}}{{\partial {\sigma ^b}}} .\label{eq:variation_induced_metric}
\end{align}
The second term can be easily calculated for polar caps, which only depends on a single parameter $b$ that is taken to be $\theta_0$.
Then, $\delta {A_b^{\rm extr}}$ follows by varying the relation \eqref{eq:unperturbed_area} with respect to the cap parameter $\theta_0$.

The individual terms $\delta {A_g^{\rm extr}}$ and $\delta {A_b^{\rm extr}}$ will be explicitly computed later for small perturbations
of the $AdS_4$ metric satisfying the linearized Einstein equations with general boundary conditions.

\subsection{Modular Hamiltonian for a Polar Cap Region}
\label{subsec:dE_formula}

Unlike the Ryu-Takayanagi formula for expressing the entanglement entropy in terms of the bulk gravitational theory,
there is no similar expression for the modular Hamiltonian. In general, the modular Hamiltonian is a non-local operator
and there is no way to this day to find an expression for the modular Hamiltonian for a general boundary state and region $A$.
In some cases, however, the modular Hamiltonian generates a geometric flow leading to a local expression.

Such an example is provided by taking a disk region $A$ of radius $R$ in a Minkowski space boundary \cite{Casini:2011kv}, in which case the
modular flow in the Cauchy development of the disk can be connected with the modular flow in Rindler space through a conformal
transformation; the details are provided in \ref{sec:modular_Hamiltonian_disc}.
Then, the modular Hamiltonian is expressed in terms of the holographic energy-momentum tensor and a conformal Killing vector
that leaves invariant the entangling curve $\partial A$ and its causal development, as
\begin{equation}
\delta E_A = \int_C {d{\Sigma ^\mu }{\mathcal{T}_{\mu \nu }}{\zeta^\nu }} ,
\label{eq:dMH_MH_Killing}
\end{equation}
where $C$ is a space-like surface with boundary $\partial A$ and ${d{\Sigma ^\mu }}$ is the differential volume form
on $C$ \cite{Wong:2013gua}.
If $C$ is selected to be a constant time slice, it will coincide with the region $A$. For a disk in Minkowski space, this
conformal Killing vector $\zeta$ is a linear combination of the Killing vectors corresponding to the time translation and a
special conformal transformation; it is selected so that $\zeta$ vanishes on $\partial A$ and its causal development.

Using the coordinate transformation \eqref{eq:coordinate_transformation}, the conformal Killing vector in question takes
the following form in global coordinates (for more details see \ref{sec:coordinate_transformation}),
\begin{equation}
\zeta = \frac{2 \pi}{{\sin {\theta _0}}}\left[ {\left( {\cos \left( {t - {t_0}} \right)\cos \theta  - \cos {\theta _0}} \right){\partial _t}
- \sin \left( {t - {t_0}} \right)\sin \theta {\partial _\theta }} \right] ,
\label{eq:dMH_Killing_vector}
\end{equation}
which indeed satisfies the conformal Killing vector equation
\begin{equation}
{\nabla _\mu }{\zeta_\nu } + {\nabla _\nu }{\zeta_\mu } - \frac{2}{d}{g_{\mu \nu }}{\nabla _\lambda }{\zeta^\lambda } = 0
\label{eq:dMH_conformal_Killing_equation}
\end{equation}
and at the same time vanishes at the entangling surface $\theta = \theta_0$ and $t=t_0$.
The conformal Killing vector $\zeta$ is the boundary limit of a conformal Killing vector field $\xi$ in the bulk,
\begin{multline}
\xi  = \frac{2 \pi}{{\sin {\theta _0}}}\left[ {\left( {\frac{{r\cos \theta \cos \left( {t - {t_0}} \right)}}{{\sqrt {{r^2} + 1} }} - \cos {\theta _0}} \right){\partial _t}} \phantom{\frac{{\sqrt {{r^2} + 1} \sin \theta \sin \left( {t - {t_0}} \right)}}{r}} \right.\\
\left. { + \sqrt {{r^2} + 1} \cos \theta \sin \left( {t - {t_0}} \right){\partial _r} - \frac{{\sqrt {{r^2} + 1} \sin \theta \sin \left( {t - {t_0}} \right)}}{r}{\partial _\theta }} \right] ,
\label{eq:dMH_MH_Killing_bulk}
\end{multline}
which also vanishes at the entangling surface.

Note in passing that if one considers generalized theories of gravity (i.e., higher derivative corrections)
an appropriate generalization of Ryu-Takayanagi conjecture will be required. Such generalizations can be obtained through the
Iyer-Wald theorem that involves bifurcate Killing horizons generated by the conformal Killing vector $\xi$ given by \eqref{eq:dMH_MH_Killing_bulk} \cite{Faulkner:2013ica}.

Formula \eqref{eq:dMH_MH_Killing} can be evaluated for any space-like surface with boundary identical to the
entangling surface. Selecting the $t=t_0$ surface and using the specific form of the Killing vector \eqref{eq:dMH_Killing_vector},
the modular Hamiltonian for the polar cap region $\theta \le \theta_0$ is written as
\begin{equation}
\delta E_A = \frac{4 \pi^2 }{{\sin {\theta _0}}}\int_0^{{\theta _0}} {d\theta \sin \theta \left( {\cos \theta  - \cos {\theta _0}} \right){\mathcal{T}_{tt}}} .
\label{eq:dMH_MH_formula}
\end{equation}
This completes the presentation of the general formulae for the quantities of interest. In section \ref{sec:EE_perturbed_AdS4_calculation}
$\delta E$ will be calculated and compared to $\delta S$ for all different kinds of $AdS_4$ perturbations satisfying general boundary
conditions.

\section{Linearized Gravity in AdS$_4$}
\label{sec:AdS4_review}

Now we come to the theory of gravitational perturbations of $AdS_4$ space-time, allowing for general
boundary conditions. It is convenient to split the perturbations into two complementary sets with opposite
parity and reduce the linearized Einstein equations to an affective \Schrodinger problem, which is
exactly solvable. Then, we write down the holographic energy-momentum tensor for all such perturbations,
which will be used later. The material we present here is based on the discussion found in \cite{Bakas:2008zg}
(but see also \cite{Bakas:2009da} for an overview of the subject) and references therein.

\subsection{Linear AdS$_4$ Perturbations}
\label{subsec:perturbations}
We consider Einstein gravity in four space-time dimensions with negative cosmological constant $\Lambda$,
\begin{equation}
{R_{\mu \nu }} = \Lambda {g_{\mu \nu }} \, .
\label{eq:Einstein_equations}
\end{equation}
$AdS_4$ is the maximally symmetric solution whose metric takes the following form in spherical coordinates,
\begin{equation}
d{s^2} =  - f\left( r \right)d{t^2} + \frac{{d{r^2}}}{{f\left( r \right)}} + {r^2}\left( {d{\theta ^2} + {{\sin }^2}\theta d{\varphi ^2}} \right) ,
\label{eq:AdS4_metric}
\end{equation}
where
\begin{equation}
f\left( r \right) = 1 - \frac{\Lambda }{3}{r^2} .
\label{eq:metric_function_f}
\end{equation}

We set the AdS scale $\sqrt{- 3 / \Lambda}$ equal to one to simplify the presentation. For later use, we define the tortoise
coordinate $x$ as
\begin{equation}
dx = \frac{{dr}}{{f\left( r \right)}} \, ,
\label{eq:tortoise_differential}
\end{equation}
which is given explicitly as
\begin{equation}
r = \tan x \, .
\label{eq:tortoise}
\end{equation}
The tortoise coordinate is an angular variable ranging from $0$ to $\pi/2$ as $r$ varies from $0$ to $\infty$.

Next, we consider linear perturbations around the $AdS_4$ metric \eqref{eq:AdS4_metric} satisfying Einstein's equations. The
gravitational perturbations fall in two complementary classes with opposite parity, namely axial and polar perturbations.
Without loss of generality we restrict attention to axially symmetric
deformations, which are described in terms of Legendre polynomials $P_l \left( \cos \theta \right)$ instead of more general
spherical harmonics $Y_l^m \left( \theta , \phi \right)$.

\subsubsection*{Axial Perturbations}
The metric perturbations of this class are parametrized by two functions ${h_0}\left( r \right)$ and ${h_1}\left( r \right)$ as follows,
\begin{multline}
d{s^2} =  - f\left( r \right)d{t^2} + \frac{{d{r^2}}}{{f\left( r \right)}} + {r^2}\left( {d{\theta ^2} + {{\sin }^2}\theta d{\varphi ^2}} \right)\\
 + 2{e^{ - i\omega t}}\sin \theta \frac{{d{P_l}\left( {\cos \theta } \right)}}{{d\theta }}\left( {{h_0}\left( r \right)dt + {h_1}\left( r \right)dr} \right)d\varphi ,
\label{eq:axial_perturbations_metric}
\end{multline}
up to reparametrizations.
It turns out that Einstein's equations are equivalent to the effective \Schrodinger problem with respect to the tortoise coordinate,
\begin{equation}
 - \frac{{{d^2}{\Psi _{{\rm{ax}}}}\left( x \right)}}{{d{x^2}}} + \frac{{l\left( {l + 1} \right)}}{{{{\sin }^2}x}}{\Psi _{{\rm{ax}}}}\left( x \right) = {\omega ^2}{\Psi _{{\rm{ax}}}}\left( x \right) ,
\label{eq:axial_effective_Schrodinger_problem}
\end{equation}
where the functions ${h_0}\left( r \right)$ and ${h_1}\left( r \right)$ are expressed in terms of the solutions as
\begin{align}
{h_0}\left( x \right) &= \frac{i}{\omega }\frac{d}{{dx}}\left( {\tan x \, {\Psi _{{\rm{ax}}}}\left( x \right)} \right) ,\label{eq:axial_perturbations_h0}\\
{h_1}\left( x \right) &= \sin x\cos x \, {\Psi _{{\rm{ax}}}}\left( x \right) . \label{eq:axial_perturbations_h1}
\end{align}

\subsubsection*{Polar Perturbations}
The metric perturbations of this class are parametrized by three functions ${H_0}\left( r \right)$, ${H_1}\left( r \right)$ and
$K\left( r \right)$ as follows,
\begin{multline}
d{s^2} =  - f\left( r \right)d{t^2} + \frac{{d{r^2}}}{{f\left( r \right)}} + {r^2}\left( {d{\theta ^2} + {{\sin }^2}\theta d{\varphi ^2}} \right)\\
 + {e^{ - i\omega t}}{P_l}\left( {\cos \theta } \right)\left[ {{H_0}\left( r \right)\left( {f\left( r \right)d{t^2} + \frac{{d{r^2}}}{{f\left( r \right)}}} \right)} \right.\\
\left. \phantom{\left( {f\left( r \right)d{t^2} + \frac{{d{r^2}}}{{f\left( r \right)}}} \right)} { + \, 2 {H_1}\left( r \right)dtdr +
K\left( r \right){r^2}\left( {d{\theta ^2} + {{\sin }^2}\theta d{\varphi ^2}} \right)} \right] ,
\label{eq:polar_perturbations_metric}
\end{multline}
up to reparametrizations.
Similarly to the previous case, Einstein's equations for polar perturbations turn out to be equivalent to the effective \Schrodinger
problem with respect to the tortoise coordinate,
\begin{equation}
 - \frac{{{d^2}{\Psi _{\rm{pol}}}\left( x \right)}}{{d{x^2}}} + \frac{{l\left( {l + 1} \right)}}{{{{\sin }^2}x}}{\Psi _{\rm{pol}}}\left( x \right) = {\omega ^2}{\Psi _{\rm{pol}}}\left( x \right) ,
 \label{eq:polar_effective_Schrodinger_problem}
\end{equation}
which is identical to the one describing the axial perturbations. The functions ${H_0}\left( r \right)$, ${H_1}\left( r \right)$ and
$K\left( r \right)$ are expressed in terms of the solutions of the effective \Schrodinger problem as
\begin{align}
{H_0}\left( x \right) &= \left( {\frac{{l\left( {l + 1} \right)}}{2}\cot x - {\omega ^2}\sin x\cos x +
{{\cos }^2 x} \, \frac{d}{{dx}}} \right){\Psi _{\rm{pol}}}\left( x \right) , \label{eq:polar_perturbations_ho}\\
{H_1}\left( x \right) &=  - i\omega \cos x\frac{d}{{dx}}\left( {\sin x \, {\Psi _{\rm{pol}}}\left( x \right)} \right) , \label{eq:polar_perturbations_h1}\\
K\left( x \right) &= \left( {\frac{{l\left( {l + 1} \right)}}{2}\cot x + \frac{d}{{dx}}} \right){\Psi _{\rm{pol}}}\left( x \right) . \label{eq:polar_perturbations_K}
\end{align}

Thus, the spectrum and the eigen-functions of the operator $-d^2/dx^2 + l(l+1)/{\rm sin}^2 x$ in the closed interval
$x \in [0, \, \pi/2]$ completely determine the gravitational perturbations of $AdS_4$ space-time.
The general solution of the linearized Einstein equations is written as linear combination of the axial and polar solutions.

\subsection{Boundary Conditions}
\label{subsec:boundary_conditions}
The solution of effective \Schrodinger problem can be expressed in terms of hypergeometric functions. The normalizable
solution, which vanishes at $r=0$, is
\begin{equation}
\Psi \left( x \right) = \cos x{\sin ^{l + 1}}x \, F\left( {\frac{1}{2}\left( {l + 2 + \omega } \right), \,
\frac{1}{2}\left( {l + 2 - \omega } \right); \, l + \frac{3}{2}; \,{{\sin }^2}x} \right) .
\end{equation}
The other independent solution of the effective \Schrodinger problem is
\begin{equation}
\Psi \left( x \right) = \frac{{\cos x}}{{{{\sin }^l}x}} \, F\left( {\frac{1}{2}\left( { - l + 1 + \omega } \right), \,
\frac{1}{2}\left( { - l + 1 - \omega } \right); \, \frac{1}{2} - l; \, {{\sin }^2}x} \right),
\end{equation}
but it diverges as $r \to 0$ and is not normalizable.

The boundary conditions as $r \to \infty$ can be systematically described by expanding the normalizable solutions in
powers of $1/r$, as
\begin{equation}
{\Psi} = {I_0} + \frac{{{I_1}}}{r} + \frac{{{I_2}}}{{{r^2}}} + \frac{{{I_3}}}{{{r^3}}} +  \ldots .
\label{eq:solution_asymptotics}
\end{equation}
Using properties of the hypergeometric functions, it turns out that $I_0$ and $I_1$ are expressed in terms of Gamma-functions as
\begin{eqnarray}
I_0 & = & \Gamma^{-1} \left({1 \over 2} (l+2+ \omega)\right) \Gamma^{-1} \left({1 \over 2}(l+2- \omega)\right) \, , \\
I_1 & = & -2 \Gamma^{-1} \left({1 \over 2} (l+1+ \omega)\right) \Gamma^{-1} \left({1 \over 2}(l+1- \omega)\right) \, .
\end{eqnarray}
The coefficients of all other terms are expressed in terms of $I_0$ and $I_1$, but the details do not really matter.
We only note here, for later use, that
\begin{equation}
{I_2} = \frac{{{I_0}}}{2}\left( {l\left( {l + 1} \right) - {\omega ^2}} \right).
\label{eq:I2_from_I0}
\end{equation}

The boundary conditions imposed as $r \to \infty$ are solely described in terms $I_0$ and $I_1$, since
\begin{equation}
I_0 = \Psi (r = \infty) \, , ~~~~~~ I_1 = - {d \Psi \over dx} (r = \infty) \, .
\end{equation}
Thus, solutions, characterized by $I_0 = 0$ obey Dirichlet boundary conditions for the effective \Schrodinger problem,
while solutions characterized by $I_1 = 0$ obey Neumann boundary conditions. Solutions with $I_0$ and $I_1$ taking more
general values correspond to more general (mixed) boundary conditions determined by the ratio $I_0 / I_1$.
In the following, we will denote by $I_k$ the coefficients of the large $r$ expansion of $\Psi (r)$ associated to axial
perturbations and $J_k$ for the polar perturbations, for distinction.

For given boundary conditions, the spectrum is discrete; for example, for Dirichlet boundary conditions we find
\begin{equation}
\omega_D = \pm \left( 2n + l + 2 \right) ,
\end{equation}
whereas for Neumann boundary conditions the spectrum is
\begin{equation}
\omega_N = \pm \left( 2n + l + 1 \right).
\end{equation}
More general boundary conditions give rise to intermediate (generally not equidistant) frequencies. In all cases, the spectrum
is implicitly determined by ratios of Gamma-functions fixed by the ratio $I_0/I_1$ or $J_0/J_1$.

\subsection{The Holographic Energy-Momentum Tensor}
\label{subsec:energy-momentum}
The energy-momentum tensor is divergent in asymptotic AdS spaces and appropriate renormalization is required to make sense of it,
using counter-terms.
Here, we summarize the results of holographic renormalization \cite{Balasubramanian:1999re,de Haro:2000xn,Skenderis:2000in,Skenderis:2002wp} for the boundary space-time metric and the corresponding
energy-momentum tensor for gravitational perturbations of $AdS_4$ space-time satisfying general boundary conditions, following \cite{Bakas:2008zg}.
Thus, we have per sector the following results, setting Newton's constant $G_N = 1$:

\subsubsection*{Axial Perturbations}
The boundary metric after conformal rescaling takes the form
\begin{equation}
d{s^2} =  - d{t^2} + \left( {d{\theta ^2} + {{\sin }^2}\theta d{\varphi ^2}} \right) + \frac{{2i{I_0}}}{\omega }{e^{ - i\omega t}}\sin \theta \frac{{d{P_l}\left( {\cos \theta } \right)}}{{d\theta }}dtd\varphi \, ,
\label{eq:boundary_metric_axial}
\end{equation}
whereas the non-vanishing components of the energy momentum tensor for small perturbations are
\begin{align}
8\pi{\mathcal{T}_{t\varphi }} &=  - \frac{i}{{2\omega }}\left( {l - 1} \right)\left( {l + 2} \right){I_1}{e^{ - i\omega t}}\sin \theta \frac{{d{P_l}\left( {\cos \theta } \right)}}{{d\theta }} \, , \\
8\pi{\mathcal{T}_{\theta \varphi }} &=  - \frac{1}{2}{I_1}{e^{ - i\omega t}}\sin \theta \left[ {l\left( {l + 1} \right){P_l}\left( {\cos \theta } \right) + 2\cot \theta \frac{{d{P_l}\left( {\cos \theta } \right)}}{{d\theta }}} \right] \, .
\end{align}

\subsubsection*{Polar Perturbations}
In this sector, the boundary metric after conformal rescaling takes the form
\begin{equation}
d{s^2} =  - d{t^2} + \left( {1 - {J_1}{e^{ - i\omega t}}{P_l}\left( {\cos \theta } \right)} \right)\left( {d{\theta ^2} +
{{\sin }^2}\theta d{\varphi ^2}} \right) ,
\label{eq:boundary_metric_polar}
\end{equation}
whereas the non-vanishing components of the energy-momentum tensor for the corresponding small perturbations are
\begin{align}
8\pi{\mathcal{T}_{tt}} &= \frac{1}{4}\left( {l - 1} \right)l\left( {l + 1} \right)\left( {l + 2} \right){J_0}{e^{ - i\omega t}}
{P_l}\left( {\cos \theta } \right) , \label{eq:polar_energy_density}\\
8\pi{\mathcal{T}_{\theta \theta }} &=  - \frac{1}{4}l\left( {l + 1} \right)\left( {1 - {\omega ^2}} \right){J_0}{e^{ - i\omega t}}
{P_l}\left( {\cos \theta } \right) \nonumber \\
 &- \frac{1}{4}\left( {l\left( {l + 1} \right) - 2{\omega ^2}} \right){J_0}{e^{ - i\omega t}}\cot \theta \frac{{d{P_l}\left( {\cos \theta } \right)}}{{d\theta }} \, , \\
8\pi{\mathcal{T}_{\varphi \varphi }} &= \frac{1}{4}l\left( {l + 1} \right)\left( {l\left( {l + 1} \right) - 1 - {\omega ^2}} \right){J_0}
{e^{ - i\omega t}}{\sin ^2}\theta {P_l}\left( {\cos \theta } \right) \nonumber \\
 &+ \frac{1}{4}\left( {l\left( {l + 1} \right) - 2{\omega ^2}} \right){J_0}{e^{ - i\omega t}}\sin \theta \cos \theta \frac{{d{P_l}\left( {\cos \theta } \right)}}{{d\theta }} \, , \\
8\pi{\mathcal{T}_{t\theta }} &= \frac{i}{4}\omega \left( {l - 1} \right)\left( {l + 2} \right){J_0}{e^{ - i\omega t}}\frac{{d{P_l}\left( {\cos \theta } \right)}}{{d\theta }} \, .
\end{align}

In all cases, the energy-momentum tensor is traceless and covariantly conserved,
\begin{equation}
{\mathcal{T}_a}^a = 0 \, , ~~~~~~~ \nabla_a \mathcal{T}^{ab} = 0
\end{equation}
with respect to the boundary metric, as required on general grounds.

Equations \eqref{eq:boundary_metric_axial} and \eqref{eq:boundary_metric_polar} imply that for axial perturbations
Dirichlet boundary conditions for the effective \Schrodinger problem correspond to Dirichlet conditions for the metric,
while for polar perturbations Neumann boundary conditions for the effective Schrodinger problem correspond to Dirichlet
conditions for the metric.
These results will be particularly useful while computing the modular Hamiltonian as an appropriately weighted integral
of the energy density for all kind of perturbations. It can already be seen that $\delta E$ always vanishes for axial perturbations,
since ${\mathcal{T}_{tt}} = 0$ irrespective
of boundary conditions, whereas $\delta E$ does not vanish for polar perturbations satisfying boundary conditions other than $J_0 = 0$.

\section{Entanglement Entropy for Gravitational Perturbations}
\label{sec:EE_perturbed_AdS4_calculation}
We are now in position to compute holographically the variation of the entanglement entropy and the modular Hamiltonian for
gravitational perturbations of $AdS_4$ space-time satisfying general boundary conditions. First, we compute the variation of
the area of the extremal surface emanating from the loop $\partial A$, as $\delta A^{\rm extr} = \delta A_g^{\rm extr} + \delta A_b^{\rm extr}$,
then we compute $\delta E$, and, finally, we compare the results with the first law of thermodynamics for the entanglement entropy,
$\delta S = \delta E$. We find that the deformations of the entangling curve $\partial A$ should be isoperimetric, when the
boundary conditions of the metric allow for fluctuating geometries on the boundary of $AdS_4$ for, otherwise, the first law of
thermodynamics would not be realized holographically. Our results generalize earlier work on the subject, going from Dirichlet to
more general boundary conditions for the metric.

\subsection{Calculation of $\delta A_g^{\rm extr}$}
\label{subsec:dAg}
First, we work out the contribution $\delta A_g^{\rm extr}$ to the variation of the area of the minimal surface.
Naturally, we consider the effect of axial and polar perturbations separately.

\subsubsection*{Axial Perturbations}
\label{subsubsec:dAg_axial}
For axial perturbations, the variation of the induced metric on the minimal surface, which is generally given by equation
\eqref{eq:variation_induced_metric}, turns out to be
\begin{align}
\delta {\gamma _{rr}} &= \delta {\gamma _{\varphi \varphi }} = 0 \, ,\\
\delta {\gamma _{r\varphi }} &= \delta {g_{r\varphi }} = {\left. {h_1}\left( r \right){e^{ - i\omega t_0}}\sin \theta \frac{\partial }{{\partial \theta }}{P_l}\left( {\cos \theta } \right) \right|}_{\theta = \arccos \left( {\cos {\theta _0}\sqrt {1 + \frac{1}{{{r^2}}}} } \right)}
\end{align}
and so the corresponding variation of the area, given by equation \eqref{eq:dAg_formula}, is
\begin{equation}
\delta {A_g^{\rm extr}} = 0 \, .
\end{equation}

\subsubsection*{Polar Perturbations}
\label{subsubsec:dAg_polar}
For polar perturbations, the variation of the induced metric \eqref{eq:variation_induced_metric} turns out to be
\begin{align}
&\delta {\gamma _{rr}} = \delta {g_{rr}} + {\left( {\frac{{d\theta }}{{dr}}} \right)^2}\delta {g_{\theta \theta }} \nonumber\\
&= \frac{1}{{1 + {r^2}}}\left( {{H_0}\left( r \right) + \frac{{{{\cos }^2}\theta_0 K\left( r \right)}}{{{r^2}\left( {1 - \left( {1 + \frac{1}{{{r^2}}}} \right){{\cos }^2}\theta_0 } \right)}}} \right){e^{ - i\omega t_0}}{P_l}\left( {\cos {\theta _0}\sqrt {1 + \frac{1}{{{r^2}}}} } \right) , \\
&\delta {\gamma _{\varphi \varphi }} = \delta {g_{\varphi \varphi }} = {r^2}\left( {1 - \left( {1 + \frac{1}{{{r^2}}}} \right){{\cos }^2}\theta_0 } \right)K\left( r \right){e^{ - i\omega t_0}}{P_l}\left( {\cos {\theta _0}\sqrt {1 + \frac{1}{{{r^2}}}} } \right) , \\
&\delta {\gamma _{r\varphi }} = 0 \, .
\end{align}

Then, using equations \eqref{eq:unperturbed_induced_metric_rr}, \eqref{eq:unperturbed_induced_metric_phiphi} and \eqref{eq:dAg_formula}, we find after some algebra that
\begin{multline}
\delta {A_g^{\rm extr}} = \pi \int_{\cot {\theta _0}}^\infty
{dr\frac{{\sin {\theta _0}}}{{\sqrt {\left( {1 + \frac{1}{{{r^2}}}} \right)} }}
\left[ {\left( {{H_0}\left( r \right) + K\left( r \right)} \right)} \phantom{\frac{1}{{{r^2}{{\tan }^2}{\theta _0}}}} \right.} \\
\left. { + \frac{1}{{{r^2}{{\tan }^2}{\theta _0}}}\left( {K\left( r \right) - {H_0}\left( r \right)} \right)} \right]
{e^{ - i\omega t_0}}{P_l}\left( {\cos {\theta _0}\sqrt {1 + \frac{1}{{{r^2}}}} } \right) .
\end{multline}
Trading $r$ with the tortoise coordinate $x$ and expressing $H_0$ and $K$ in terms of the $\Psi_{\rm pol}$, as given by equations \eqref{eq:polar_perturbations_ho} and \eqref{eq:polar_perturbations_K}, we obtain
\begin{multline}
\delta {A_g^{\rm extr}} = \pi \int_{\frac{\pi }{2} - {\theta _0}}^{\frac{\pi }{2}} {dx\frac{{\sin {\theta _0}}}{{\cos x}}\left[ {l\left( {l + 1} \right){\Psi _{\rm{pol}}} - { \left( {1 - \frac{{{{\cos }^2}x}}{{{{\sin }^2}{\theta _0}}}} \right)}{\omega ^2}{\Psi _{\rm{pol}}}} \right.} \\
\left. { + \tan x\left( {1 + \frac{{{{\cos }^2}x}}{{{{\sin }^2}{\theta _0}}}} \right)\frac{{d{\Psi _{\rm{pol}}}}}{{dx}}} \right] {e^{ - i\omega t_0}}{P_l}\left( {\frac{{\cos {\theta _0}}}{{\sin x}}} \right) .
\end{multline}
The effective \Schrodinger problem \eqref{eq:polar_effective_Schrodinger_problem} can be used further to eliminate $\omega$, leading
to the following expression,
\begin{multline}
\delta {A_g^{\rm extr}} = \pi \int_{\frac{\pi }{2} - {\theta _0}}^{\frac{\pi }{2}} {dx\left[ {\frac{d}{{dx}}\left[ {\left( {\frac{{\sin {\theta _0}}}{{\cos x}} - \frac{{\cos x}}{{\sin {\theta _0}}}} \right)\frac{{d{\Psi _{\rm{pol}}}}}{{dx}}} \right]} \right.} \\
\left. { + l\left( {l + 1} \right)\frac{{\cos {\theta _0}\cot x}}{{\tan {\theta _0}\sin x}} \,
{\Psi _{\rm{pol}}}} \right]{e^{ - i\omega t_0}}{P_l}\left( {\frac{{\cos {\theta _0}}}{{\sin x}}} \right).
\end{multline}
It is convenient at this point to employ the following identities among Legendre polynomials,
\begin{align}
\frac{d}{{d \xi}}\left( {{P_{l + 1}}\left( \xi \right) - {P_{l - 1}}\left( \xi \right)} \right) &=
\left( {2l + 1} \right){P_l}\left( \xi \right) , \label{eq:dAg_Legendre_identity_1}\\
\frac{d}{{d \xi}}{P_l}\left( \xi \right) &= \frac{{l\left( {l + 1} \right)}}{{2l + 1}}\frac{1}{{{\xi^2} - 1}}
\left( {{P_{l + 1}}\left( \xi \right) - {P_{l - 1}}\left( \xi \right)} \right) ,\label{eq:dAg_Legendre_identity_2}
\end{align}
to perform integration by parts and get
\begin{multline}
\delta {A_g^{\rm extr}} = \pi {e^{ - i\omega t_0}} \left[ {\left( {\frac{{\sin {\theta _0}}}{{\cos x}} -
\frac{{\cos x}}{{\sin {\theta _0}}}} \right)\frac{{d{\Psi _{\rm{pol}}}}}{{dx}} \,
{P_l}\left( {\frac{{\cos {\theta _0}}}{{\sin x}}} \right) } \right]_{\frac{\pi }{2} - {\theta _0}}^{\frac{\pi }{2}}\\
 - \pi {e^{ - i\omega t_0}} \left[ {\cot {\theta _0} \, \frac{{l\left( {l + 1} \right)}}{{2l + 1}}
 \left( {{P_{l + 1}}\left( {\frac{{\cos {\theta _0}}}{{\sin x}}} \right) -
 {P_{l - 1}}\left( {\frac{{\cos {\theta _0}}}{{\sin x}}} \right)} \right)} {\Psi _{\rm{pol}}}
 \right]_{\frac{\pi }{2} - {\theta _0}}^{\frac{\pi }{2}} .
\end{multline}
The contribution from the end-point $x = \frac{\pi}{2} - \theta_0$ vanishes for both bracketed terms, and, thus, the result is
\begin{multline}
\delta {A_g^{\rm extr}} = \pi {e^{ - i\omega t_0}} \mathop {\lim }\limits_{x \to \frac{\pi }{2}}
\left[ {\frac{{\sin {\theta _0}}}{{\cos x}}\frac{{d{\Psi _{\rm{pol}}}}}{{dx}} \, {P_l}\left( {\cos {\theta _0}} \right) }\right. \\
\left. {- \cot {\theta _0} \, \frac{{l\left( {l + 1} \right)}}{{2l + 1}} \left( {{P_{l + 1}}\left( {\cos {\theta _0}} \right) -
{P_{l - 1}}\left( {\cos {\theta _0}} \right)} \right)} \, {\Psi _{\rm{pol}}} \right] .
\label{eq:dAg_formula_limit_x}
\end{multline}

Going back to the coordinate $r$ and using the asymptotic expansion of the \Schrodinger wave-functions \eqref{eq:solution_asymptotics}
together with equation \eqref{eq:I2_from_I0}, we arrive at the final expression
\begin{multline}
\delta {A_g^{\rm extr}} =  - \pi {e^{ - i\omega t_0}}\left[ {\mathop {\lim }\limits_{r \to \infty }
\left( {r{J_1} + {J_0}\left( {l\left( {l + 1} \right) - {\omega ^2}} \right)} \right) \sin {\theta _0} \,
{P_l}\left( {\cos {\theta _0}} \right)} \phantom{\frac{{l\left( {l + 1} \right)}}{{2l + 1}}} \right.\\
\left. { + \cot {\theta _0} \, \frac{{l\left( {l + 1} \right)}}{{2l + 1}}{J_0}
\left( {{P_{l + 1}}\left( {\cos {\theta _0}} \right) - {P_{l - 1}}\left( {\cos {\theta _0}} \right)} \right)} \right].
\label{eq:dAg_formula_limit_r}
\end{multline}
Note that $\delta {A_g^{\rm extr}}$ is divergent when the perturbations of the metric do not satisfy Dirichlet
boundary conditions, $J_1 \neq 0$.

\subsection{Calculation of $\delta A_b^{\rm extr}$}
\label{subsec:dAb}
Since we are considering axially symmetric perturbations of the bulk metric, the deformations of the entangling curve
are taken to be independent of the angular coordinate $\phi$ to conform with the symmetry. Thus, the variations of
the parameter $\theta_0$ in the polar cap are taken to be independent of $\phi$, i.e.,
\begin{equation}
\theta_0 \left( t \right) = \theta_0 + \delta \theta \left( t \right) .
\end{equation}
Consequently, $\delta {A_b^{\rm extr}}$ can be derived directly by varying \eqref{eq:unperturbed_area} with respect to $\theta_0$,
leading to the following result
\begin{equation}
\delta {A_b^{\rm extr}} = 2\pi \mathop {\lim }\limits_{r \to \infty } r\cos {\theta _0} \, \delta \theta \left( t_0 \right)
\end{equation}
for all kind of metric perturbations.

Note that $\delta A_b^{\rm extr}$ also appears to be divergent on general grounds. It is important to see how these
divergences combine with those appearing in $\delta A_g^{\rm extr}$ to form $\delta A^{\rm extr}$.

\subsection{Net Variation of the Entanglement Entropy}
\label{subsec:dA}
Summing up the two contributions, we obtain the net variation of the area of the minimal surface, $\delta A^{\rm extr}$,
and, hence, the variation of entanglement entropy, $\delta S_A$. We present the result for axial and polar perturbations,
\begin{flushleft}
setting Newton's constant $G_N = 1$. Thus, for axial perturbations, we have
\begin{equation}
\delta S_A^{{\rm{axial}}} = \frac{\pi }{2}\mathop {\lim }\limits_{r \to \infty } r\cos {\theta _0} \,
\delta {\theta}^{{\rm{axial}}}\left( t_0 \right) ,
\label{eq:entanglement_entropy_result_axial}
\end{equation}
\end{flushleft}
whereas for polar perturbations the result reads
\begin{multline}
\delta S_A^{{\rm{polar}}} =  - \frac{\pi }{4}{e^{ - i\omega t_0}}\left[ {\mathop {\lim }\limits_{r \to \infty }
\left( {r{J_1} + {J_0}\left( {l\left( {l + 1} \right) - {\omega ^2}} \right)} \right) \sin {\theta _0} \,
{P_l}\left( {\cos {\theta _0}} \right)} \phantom{\frac{{l\left( {l + 1} \right)}}{{2l + 1}}} \right.\\
\left. { + \cot {\theta _0} \, \frac{{l\left( {l + 1} \right)}}{{2l + 1}}{J_0}
\left( {{P_{l + 1}}\left( {\cos {\theta _0}} \right) - {P_{l - 1}}\left( {\cos {\theta _0}} \right)} \right)} \right] \\
 + \frac{\pi }{2}\mathop {\lim }\limits_{r \to \infty } r\cos {\theta _0} \, \delta {\theta}^{{\rm{polar}}}\left( t_0 \right) .
\label{eq:entanglement_entropy_result_polar}
\end{multline}

The variance of the entropy appears to be divergent, which is unphysical. Specific conditions should be imposed
on $\delta {\theta}^{{\rm{polar}}}(t_0)$ to make them cancel, but these will be discussed later, while comparing $\delta S_A$
with $\delta E_A$, together with their geometric meaning.

\subsection{Modular Hamiltonian for Gravitational Perturbations}
\label{sec:MH_perturbed_AdS4_calculation}
Next, we compute the variation of the modular Hamiltonian by specializing formula \eqref{eq:dMH_MH_formula} to axial and
polar perturbations satisfying general boundary conditions. As before, Newton's constant is set equal
to 1.

\subsubsection*{Axial Perturbations}
\label{subsubsec:dE_axial}

Since $\mathcal{T}_{tt} = 0$ for the axial perturbations, we immediately obtain
\begin{equation}
\delta E_A^{{\rm{axial}}} = 0
\label{eq:modular_Hamiltonian_result_axial}
\end{equation}
in all cases, irrespective of boundary conditions.

\subsubsection*{Polar Perturbations}
\label{subsubsec:dE_polar}
For polar perturbations with $J_0 \neq 0$, the component $\mathcal{T}_{tt}$ of the holographic energy-momentum tensor
does not vanish, and, hence, the corresponding variance of the modular Hamiltonian differs from zero.

First, substituting the expression \eqref{eq:polar_energy_density} for $\mathcal{T}_{tt}$ into equation \eqref{eq:dMH_MH_formula},
yields the following expression for the variance of the modular Hamiltonian,
\begin{equation}
\delta E_A^{{\rm{polar}}} = \frac{\pi \left( {l - 1} \right)l\left( {l + 1} \right)\left( {l + 2} \right)}{{8 \sin {\theta _0}}}{J_0}{e^{ - i\omega {t_0}}}\int_0^{{\theta _0}} {d\theta \sin \theta \left( {\cos \theta  - \cos {\theta _0}} \right){P_l}\left( {\cos \theta } \right)} \, .
\end{equation}
Then, we perform integration by parts and make use of formula \eqref{eq:dAg_Legendre_identity_1} twice to find
\begin{multline}
\delta E_A^{{\rm{polar}}} = \frac{\pi }{{8 \sin {\theta _0}}}\frac{{\left( {l - 1} \right)l\left( {l + 1} \right)\left( {l + 2} \right)}}{{2l + 1}}{J_0}{e^{ - i\omega {t_0}}}\\
\times \left[ {\frac{{{P_{l + 2}}\left( {\cos {\theta _0}} \right)}}{{2l + 3}} + \frac{{{P_{l - 2}}\left( {\cos {\theta _0}} \right)}}{{2l - 1}} - \frac{{2\left( {2l + 1} \right){P_l}\left( {\cos {\theta _0}} \right)}}{{\left( {2l - 1} \right)\left( {2l + 3} \right)}}} \right] .
\end{multline}

Next, using the following identity of Legendre polynomials,
\begin{equation}
\left( {l + 1} \right){P_{l + 1}}\left( \xi \right) + l{P_{l - 1}}\left( \xi \right) = \left( {2l + 1} \right)\xi \,
{P_l}\left( \xi \right)
\label{eq:dMH_Legendre_identity_1}
\end{equation}
to eliminate $P_{l-2}$ and $P_{l+2}$, we arrive at the expression
\begin{multline}
\delta E_A^{{\rm{polar}}} = \frac{\pi }{{8 \sin {\theta _0}}}\frac{{\left( {l - 1} \right)l\left( {l + 1} \right)\left( {l + 2} \right)}}{{2l + 1}}{J_0}{e^{ - i\omega {t_0}}} \\
\times \left[ {\cos {\theta _0}\left( {\frac{{{P_{l + 1}}\left( {\cos {\theta _0}} \right)}}{{l + 2}} + \frac{{{P_{l - 1}}\left( {\cos {\theta _0}} \right)}}{{l - 1}}} \right) - \frac{{\left( {2l + 1} \right){P_l}\left( {\cos {\theta _0}} \right)}}{{\left( {l - 1} \right)\left( {l + 2} \right)}}} \right] ,
\end{multline}
which can be conveniently rewritten as
\begin{multline}
\delta E_A^{{\rm{polar}}} = \frac{\pi }{{8 \sin {\theta _0}}}\frac{{\left( {l - 1} \right)l\left( {l + 1} \right)\left( {l + 2} \right)}}{{2l + 1}}{J_0}{e^{ - i\omega {t_0}}} \\
\times \left[ {\frac{{\cos {\theta _0}}}{{\left( {l - 1} \right)\left( {l + 2} \right)}}\left( {\left( {l + 1} \right){P_{l + 1}}\left( {\cos {\theta _0}} \right) + l{P_{l - 1}}\left( {\cos {\theta _0}} \right)} \right)} \right.\\
\left. { - \frac{{2\cos {\theta _0}}}{{\left( {l - 1} \right)\left( {l + 2} \right)}}\left( {{P_{l + 1}}\left( {\cos {\theta _0}} \right) - {P_{l - 1}}\left( {\cos {\theta _0}} \right)} \right) - \frac{{\left( {2l + 1} \right){P_l}\left( {\cos {\theta _0}} \right)}}{{\left( {l - 1} \right)\left( {l + 2} \right)}}} \right] .
\end{multline}

Using once more the identity \eqref{eq:dMH_Legendre_identity_1} yields the final result,
\begin{multline}
\delta E_A^{{\rm{polar}}} =  - \frac{\pi }{{4}}{e^{ - i\omega t_0}}\left[ {\frac{1}{2}{J_0} \,
l\left( {l + 1} \right)\sin {\theta _0}{P_l}\left( {\cos {\theta _0}} \right)} \right.\\
\left. { + \cot {\theta _0} \, \frac{{l\left( {l + 1} \right)}}{{2l + 1}}{J_0}\left( {{P_{l + 1}}\left( {\cos {\theta _0}} \right) -
{P_{l - 1}}\left( {\cos {\theta _0}} \right)} \right)} \right] .
\label{eq:modular_Hamiltonian_result_polar}
\end{multline}

We see that in all cases the variance of the modular Hamiltonian is finite.

\subsection{Comparison with First Law of Thermodynamics}
The first law of thermodynamics $\delta S = \delta E$ for the entanglement entropy and the modular Hamiltonian is a tautology
from the point of view of the boundary theory. Its realization through the Ryu-Takayanagi formula \eqref{eq:RT_conjecture} and
the holographic energy-momentum tensor is not automatic, but it is a consistency check of the whole scheme. Comparing
the results derived above, we find that $\delta S_A = \delta E_A$ for any polar cap region A, provided that the entangling curve
at the boundary also deforms, where appropriate, in a certain way. The boundary is not held fixed when general boundary conditions
are imposed on the metric, and, thus, the polar cap region $A$ and the entangling curve $\partial A$ also vary. Thus, $\delta \theta (t_0)$
should be adjusted accordingly to achieve agreement with the first law of thermodynamics. As will be seen shortly this requirement has
natural geometric interpretation. It also gets rid of the divergences appearing in the computation of $\delta S_A$.

\subsubsection*{Axial Perturbations}
\label{subsubsec:dAb_axial}
First, we consider the case of axial perturbations and compare the expressions \eqref{eq:entanglement_entropy_result_axial} for
$\delta S_A$ and \eqref{eq:modular_Hamiltonian_result_axial} for $\delta E_A$. The first law is realized holographically provided that
\begin{equation}
\delta {\theta}^{{\rm{axial}}}\left( t_0 \right) = 0
\label{eq:delta_theta_0_axial}
\end{equation}
irrespective of boundary conditions. Thus, axial perturbations do not require adjustment of the polar cap region.

A way to understand this is to consider the boundary metric of the perturbed $AdS_4$ space-time, prior to conformal scaling,
and work out the two-dimensional induced metric on the corresponding polar cap region. For axial perturbations it reads
\begin{equation}
{ds^2}_{\rm{axial}} = {r^2}\left( {d{\theta ^2} + {{\sin }^2}\theta d{\varphi ^2}} \right) .
\end{equation}
The polar cap is placed on a round sphere that remains inert to perturbations irrespective of boundary conditions.

\subsubsection*{Polar Perturbations}
\label{subsubsec:dAb_polar}
Next, we consider the case of polar perturbations and compare the corresponding expressions \eqref{eq:entanglement_entropy_result_polar}
for $\delta S_A$ and \eqref{eq:modular_Hamiltonian_result_polar} for $\delta E_A$.
The first law of thermodynamics for entanglement entropy is realized holographically provided that the entangling curve in
the polar cap region varies according to the equation
\begin{equation}
\delta {\theta}^{{\rm{polar}}}\left( t_0 \right) = \frac{1}{2} {e^{ - i\omega t_0}}\left[ {{J_1} +
\frac{{J_0}}{{r}}\left( {\frac{1}{2} l\left( {l + 1} \right) - {\omega ^2}} \right)} \right]\tan {\theta _0} \,
{P_l}\left( {\cos {\theta _0}} \right) .
\label{eq:delta_theta_0_polar}
\end{equation}
This also takes care of the unphysical divergences appearing in $\delta S_A$.
Note at this end that a subleading term proportional to $J_0$ is included in the variation
$\delta {\theta}^{{\rm{polar}}}(t_0)$; it contributes a finite term to $\delta S_A$ which should also be
removed to match it with $\delta E_A$.

In this case, perturbations of the bulk metric require adjusting the entangling curve in the boundary.
To understand what is happening, let us consider the boundary metric of the perturbed $AdS_4$ space-time, prior to conformal scaling,
and work out the two-dimensional induced metric on the corresponding polar cap region $A$. For polar perturbations, the
induced metric on the spherical slices of the boundary is
\begin{equation}
{ds^2}_{\rm{polar}} = {r^2}\left( {1 + K\left( r \right){e^{ - i\omega t}}{P_l}\left( {\cos {\theta}} \right)} \right)
\left( {d{\theta ^2} + {{\sin }^2}\theta d{\varphi ^2}} \right) ,
\label{eq:spherical_region_metric}
\end{equation}
where $K(r)$ is the radial function appearing in the general parametrization \eqref{eq:polar_perturbations_metric} of the bulk space metric.
The asymptotic form for $K\left( r \right)$ follows from the large $r$ expansion of the \Schrodinger wave function $\Psi_{\rm pol}$,
via equation \eqref{eq:polar_perturbations_K}, and it reads
\begin{equation}
K\left( r \right) =  - {J_1} - \frac{{J_0}}{{r}}\left( {\frac{1}{2} l\left( {l + 1} \right) - {\omega ^2}} \right) + \mathcal{O}\left( {\frac{1}{{{r^2}}}} \right) .
\end{equation}
It implies, in particular, that the condition \eqref{eq:delta_theta_0_polar} on $\delta {\theta}^{{\rm{polar}}}\left( t_0 \right)$ takes the following
equivalent form
\begin{equation}
\delta {\theta}^{{\rm{polar}}}\left( t_0 \right) = - \frac{1}{2} {e^{ - i\omega t_0}} K(r) \tan {\theta _0} \,
{P_l}\left( {\cos {\theta _0}} \right)
\label{eq:our_guy}
\end{equation}
up to irrelevant subleading terms.

The deformation of the entangling curve is such that its line element is inert to the perturbation. Using the spherical metric \eqref{eq:spherical_region_metric}, we find that
the induced line element of the boundary curve of a polar cap region with parameter $\theta$ is
\begin{equation}
dl = \mathop {\lim }\limits_{r \to \infty } \sqrt {{g_{\varphi \varphi }}} \, d\varphi  = \mathop {\lim }\limits_{r \to \infty } \left( {1 + \frac{1}{2}K\left( r \right){e^{ - i\omega t}}{P_l}\left( {\cos \theta } \right)} \right)r\sin \theta \, d\varphi \, .
\end{equation}
Here, we also take the large $r$ limit as the computations take place on the boundary of space-time, prior to rescaling.
Setting $\theta = \theta_0 + \delta \theta (t_0)$, where $\delta \theta (t_0)$ is provided by \eqref{eq:our_guy}, we have to
linear order that
\begin{eqnarray}
&\left( {1 + \frac{1}{2} K\left( r \right){e^{ - i\omega t_0}}{P_l}\left( {\cos ({\theta _0}} + \delta \theta \left(t_0\right))\right)} \right)r
\sin \left( {{\theta _0} + \delta {\theta}\left( t_0 \right)} \right) = &\nonumber \\
&\left( {1 + \frac{1}{2} K\left( r \right){e^{ - i\omega t_0}}{P_l}\left( {\cos {\theta _0}} \right)} \right)r\sin \left( {{\theta _0} + \delta {\theta}\left( t_0 \right)} \right) = r\sin {\theta _0} +\mathcal{O} \left( \frac{1}{r} \right) .&
\end{eqnarray}
Thus, although the polar cap region and its boundary undergo perturbations, the line element of the entangling curve remains
constant,
\begin{equation}
dl = \mathop {\lim }\limits_{r \to \infty } r\sin \theta_0 \, d\varphi \, .
\end{equation}

Summarizing, we see that for all kind of gravitational perturbations of $AdS_4$ space-time (axial or polar) and for all kind of boundary
conditions, the holographic realization of the first law of thermodynamics requires that the line element of the entangling curve remains invariant.
Region $A$, whose entanglement entropy is provided by formula \eqref{eq:RT_conjecture}, deforms with time in such a
way that the line element of the entangling curve remains constant.
This requirement also takes care of the divergencies that otherwise would have inflicted the variance of the entanglement entropy.
Since the first law of thermodynamics is trivially valid from the CFT point of view, the above statement should be viewed as addendum to
Ryu-Takayanagi prescription.

Although the laws of entanglement thermodynamics are not yet completely understood, it has been shown that for spherical regions the
reduced density matrix $\rho_A$ is identical to a thermal density matrix with effective inverse temperature equal to the perimeter
of the entangling circular curve \cite{Casini:2011kv}. Through this connection, our results can be thought to provide the holographic interpretation of an
isothermal process of entanglement thermodynamics.

An interesting question arises by comparing our calculation to previous ones \cite{Lashkari:2013koa,Faulkner:2013ica}. In those papers,
the first law of thermodynamics for entanglement was realized holographically using gravitational perturbations satisfying Dirichlet
boundary conditions,
without making reference to the contribution $\delta A_b^{\rm extr}$. In our case, even for Dirichlet boundary conditions for the polar
perturbations of the metric, $J_1 = 0$, $\delta A_b^{\rm extr}$ is non-vanishing. Its contribution is absolutely necessary for
validating the first law of thermodynamics for entanglement. The resolution to this ostensible contradiction is that each one
of the terms arising in $\delta A_g^{\rm extr}$ and $\delta A_b^{\rm extr}$ depend on the choice of coordinates, but their net sum
is, in fact, coordinate independent when the line element of the entangling curve remains constant.
The variation $\delta A_b^{\rm extr}$ vanishes for Dirichlet boundary conditions only in some coordinate systems, which include
the Fefferman-Graham coordinates used in references \cite{Lashkari:2013koa,Faulkner:2013ica}. More technical details about this
issue are provided in \ref{sec:coordinate_dependence}.

The demand for isometric deformations of the entangling curve is also strongly connected with renormalization issues of the entanglement entropy. In odd dimensional conformal field theories, as in our case, the finite contribution to the entanglement entropy can be contaminated by the divergent terms, using alternative definitions of the UV cutoff scale $L$. This issue has already been noticed in the literature \cite{Casini:2011kv}. The universal constant term can be isolated using an appropriate renormalized entanglement entropy, which in $2+1$ conformal field theories (and at least for special choices of the entangling curve) takes the form \cite{Casini:2012ei}
\begin{equation}
{\tilde S_{EE}}: = {S_{EE}} - \frac{{\partial {S_{EE}}}}{{\partial \, {\rm{L}}\left( {\partial A} \right)}} \, {\rm{L}}\left( {\partial A} \right) ,
\label{eq:renormalized_See}
\end{equation}
where ${\rm{L}}\left( {\partial A} \right)$ is the length of the entangling curve, resolving the above issue. This is also in agreement, via the Ryu-Takayanagi conjecture, with the definition of renormalized area in the bulk theory, provided by Graham and Witten \cite{Graham:1999pm}; in that work the same contamination arises from the divergent terms, resulting in a specific choice of radial coordinate to cure the problem. When the metric perturbations of $AdS_4$ are taken in the form \eqref{eq:axial_perturbations_metric} or \eqref{eq:polar_perturbations_metric}, the
corresponding radial coordinate $r$ does not belong to that special class, calling for a renormalized definition of holographic entanglement entropy, as in equation \eqref{eq:renormalized_See}. In this language, the term $\delta A_g^{\rm extr}$ corresponds to the variation of $S_{EE}$, whereas the term $\delta A_b^{\rm extr}$ corresponds to the variation of the counter-term. Then, as it is implied by our results, the inclusion of the counter-term is equivalent to the statement that the unrenormalized entanglement entropy, which is provided by the Ryu-Takayanagi formula \eqref{eq:RT_conjecture}, is computed for a region $A$ bounded by an entangling curve that deforms isometrically.

\section{Electric-Magnetic Duality for Entanglement Entropy}
\label{sec:duality}
In this section, we review briefly the electric-magnetic duality exhibited by linearized gravity, together with its
holographic manifestation, following \cite{Bakas:2008zg}, and examine the implications to entanglement entropy.

\subsection{Energy-Momentum/Cotton Tensor Duality}
\label{subsec:duality}

Linearized gravity around maximally symmetric backgrounds exhibits a rank-2 generalization of electric-magnetic duality
in four space-time dimensions \cite{Henneaux:2004jw,julia}. It is best described in terms of the Weyl tensor
$C_{\mu \nu \rho \sigma }$ and its dual
counterpart,
\begin{equation}
{{\tilde C}_{\mu \nu \rho \sigma }} = \frac{1}{2}{\varepsilon _{\mu \nu }}^{\kappa \lambda }{C_{\kappa \lambda \rho \sigma }},
\label{eq:duality_dual_Weyl}
\end{equation}
as follows:

The Weyl tensor vanishes identically for maximally symmetric backgrounds, $Mink_4$, $dS_4$ or $AdS_4$, which are solutions
of the full non-linear Einstein equations with cosmological constant $\Lambda$. Linearized gravity around the maximally
symmetric solution, whose choice depends upon the sign of the $\Lambda$, exhibits a symmetry under the interchange of the
electric and magnetic components of the corresponding Weyl tensor. At the level of equations, it amounts to exchanging the
role of the linearized field equations and the Bianchi identities. Two metric perturbations
$\delta \tilde{g}_{\mu \nu}$ and $\delta g_{\mu \nu}$ are said to be dual to each other if
\begin{equation}
{{\tilde C}_{\mu \nu \rho \sigma }}\left( g \right) = {C_{\mu \nu \rho \sigma }}\left( {\tilde g} \right) ,
\label{eq:duality_Weyl}
\end{equation}
so that the electric components of the one Weyl tensor are the magnetic components of the other and vice-versa. Although these relations
are highly non-local, they are naturally resolved by decomposing the gravitational perturbations into two distinct classes
of opposite parity, namely axial and polar perturbations. It turns out that
\begin{align}
{\delta {\tilde g}^{{\rm{polar}}}} = {\delta g^{{\rm{axial}}}} \, , ~~~~~~ {\delta {\tilde g}^{{\rm{axial}}}} = {\delta g^{{\rm{polar}}}} \, ,
\label{eq:duality_dual_metric_2}
\end{align}
provided that the solutions of the corresponding effective \Schrodinger problems are interrelated as
\begin{equation}
{{\tilde \Psi }_{\rm{pol}}} =  - \frac{{2i}}{\omega }{\Psi _{{\rm{ax}}}} \, , ~~~~~~
{{\tilde \Psi }_{{\rm{ax}}}} = \frac{{i\omega }}{2}{\Psi _{\rm{pol}}} \, .
\end{equation}
Since the \Schrodinger problems of the axial and polar perturbations are identical, though the precise form of the
effective potential depends on the cosmological constant $\Lambda$, the relation among $\tilde \Psi$ and $\Psi$
enforces the two classes of perturbations to have the same frequency $\omega$ and the same boundary conditions at spatial
infinity.

Specializing the discussion to gravitational perturbations of $AdS_4$ space-time, which is the subject of the present work,
one finds that electric-magnetic duality has a holographic manifestation \cite{Leigh:2007wf,deHaro:2008gp,Bakas:2008zg,Bakas:2009da}.
Quite remarkably, the holographic energy-momentum
tensor $\mathcal{T}_{ab}$ and the Cotton tensor $\mathcal{C}_{ab}$ of the metric at the boundary of space-time follow
from the electric and magnetic components of the bulk Weyl tensor (with respect to the radial ADM decomposition of the
metric) as
\begin{align}
8\pi {\mathcal{T}_{ab}} &=  - \mathop {\lim }\limits_{r \to \infty } {r^3}{C_{a r b r}} \, , \label{eq:duality_stress_energy}\\
{\mathcal{C}_{ab}} &= \mathop {\lim }\limits_{r \to \infty } {r^3}{{\tilde C}_{a r b r}} \, . \label{eq:duality_Cotton}
\end{align}
The Cotton tensor is traceless, symmetric and covariantly conserved, as for the energy-momentum tensor, and it measures the
deviations of the boundary metric from conformal flatness due to the boundary conditions one imposes on the perturbations
of the metric.
The last equations are general and applicable to the full non-linear regime of $AdS_4$ gravity, but when specialized to
the linearized theory they give rise to the so called energy-momentum/Cotton tensor duality that is resolved in sectors as
\begin{align}
8\pi {\mathcal{T}_{ab}^{{\rm{polar}}}} = {\mathcal{C}_{ab}^{{\rm{axial}}}}, \label{eq:duality1}\\
8\pi {\mathcal{T}_{ab}^{{\rm{axial}}}} = {\mathcal{C}_{ab}^{{\rm{polar}}}}. \label{eq:duality2}
\end{align}

\subsection{Duality for Entanglement Entropy}
The extremal surface $A^{\rm extr}$, whose area provides the entanglement entropy, is defined with respect to a given bulk metric
$g$ for a given boundary region. Let us define ${{\tilde A}^{{\rm{extr}}}}$ as the extremal surface taken with respect to the dual
metric ${\tilde g}$, following \eqref{eq:duality_Weyl},
\begin{equation}
{{\tilde A}^{{\rm{extr}}}}\left( g \right) = {A^{{\rm{extr}}}}\left( {\tilde g} \right) .
\end{equation}
The corresponding entanglement entropy is taken to be
\begin{equation}
{{\tilde S}_A} = {1 \over 4} \, {\rm{Area}}\left( {{{\tilde A}^{{\rm{extr}}}}} \right) \, .
\end{equation}

Since the unperturbed $AdS_4$ space is self dual, ${{\tilde g}_{\rm AdS}} = {g_{\rm AdS}}$, the unperturbed extremal
surface is also self-dual. The variation of the dual entanglement entropy for axial and polar perturbations, which are dual to each other,
can be calculated in the same way as the variation of entanglement entropy. As a direct consequence of equations
\eqref{eq:duality_dual_metric_2}, we obtain
\begin{align}
\delta {{\tilde S}^{{\rm{axial}}}} &= \delta {S^{{\rm{polar}}}} ,\\
\delta {{\tilde S}^{{\rm{polar}}}} &= \delta {S^{{\rm{axial}}}} .
\end{align}

Similarly, one can define the variation of the dual modular Hamiltonian using the Cotton tensor of the original boundary metric, as
\begin{equation}
\delta \tilde E = \frac{1}{8\pi}\int_C {d{\Sigma ^\mu }{\mathcal{C}_{\mu \nu }}{\zeta^\nu }} .
\end{equation}
As direct consequence of equations \eqref{eq:duality1} and \eqref{eq:duality2}, it turns out that
\begin{align}
\delta {{\tilde E}^{{\rm{axial}}}} &= \delta {E^{{\rm{polar}}}} ,\\
\delta {{\tilde E}^{{\rm{polar}}}} &= \delta {E^{{\rm{axial}}}} .
\end{align}

It is true that $\delta S = \delta E$ for all kind of perturbations, and, thus, it is also true that
\begin{equation}
\delta \tilde S = \delta \tilde E \, ,
\end{equation}
showing that the first law of thermodynamics for entanglement is self-dual under gravitational duality.

Note in this context that the boundary metric is not identical to the dual boundary metric, since the duality transformation
interchanges Dirichlet and Neumann boundary conditions for the metric (it only preserves the boundary conditions of the
effective \Schrodinger problems). Since the line element of the entangling curve remains inert to perturbations, the entangling
curve is not identical to the dual entangling curve. Likewise, the polar cap region $A$ is not the same as its dual counterpart $\tilde{A}$. These differences should be taken into account in the definitions of the dual thermodynamic quantities above.

\section{Conclusions and Discussion}
\label{sec:conclusions}
In this paper we studied the variation of the entanglement entropy and the expectation value of the modular Hamiltonian for
small gravitational perturbations of $AdS_4$ space-time satisfying general boundary conditions. We found that the first law of
thermodynamics for entanglement is realized holographically through the Ryu-Takayanagi formula, provided that the line element
of the entangling curve on the boundary remains constant, hereby generalizing previous results on the subject. It should be viewed as addendum to the Ryu-Takayanagi prescription.
In this context, we also noted that the perimeter of the entangling curve is the ``area'' appearing in the ``area law'' term for
entanglement entropy. Thus, our demand that the line element of the entangling curve remains constant is equivalent to the statement
that the ``area law'' term is inert to metric perturbations for all different kind of boundary conditions. In turn, this appears to be equivalent to using a certain renormalized form for the holographic entanglement entropy.

Although our presentation
is technically limited to metric perturbations and entangling curves that are axially symmetric, choosing polar cap regions with circular
boundary, we expect it to generalize in the absence of axial symmetry. It will be interesting to see how exactly this happens.

We also examined the implications of electric-magnetic duality of linearized gravity for the entanglement entropy and the
associated first law of thermodynamics. We defined a dual entanglement entropy and a dual modular Hamiltonian and found that they
also satisfy the first law of thermodynamics for appropriately related dual regions $A$ and $\tilde{A}$ at the boundary of $AdS_4$.
These results too can be thought to provide an additional consistency check of the Ryu-Takayanagi conjecture.

Typically, electric-magnetic duality connects the dual degrees of freedom of the theory in a non-trivial and non-local way.
In the absence of excitations, it reduces to a trivial local relation, but this is not any more so in the presence of excitations.
If one defines the entanglement entropy in a region by tracing out the degrees of freedom in the complement,
the same degrees of freedom will correspond to a different region in the dual description. Thus, the dual
entanglement entropy should be defined for a different entangling curve.
Our results, showing that the holographic realization of the first law of thermodynamics is achieved only when
the line element of the entangling curve remains constant, are in accord with the different boundary conditions satisfied by
dual gravitational excitations so that the dual boundary metrics differ.

It would be interesting to see if such considerations can also be applied to other backgrounds such as $AdS_4$ black holes. Perturbations
of black-holes exhibit a duality relation which is best described in terms of supersymmetric partner potentials for the
effective \Schrodinger problems of the axial and polar sectors. Although this duality has no direct interpretation in
terms of the electric and magnetic components of the Weyl tensor in the bulk, as in $AdS_4$ space-time, it has holographic manifestation
as energy-momentum/Cotton tensor duality for the hydrodynamic modes of black holes \cite{Bakas:2008gz}. We hope to address
the related thermal effects of black holes elsewhere.

\section*{Acknowledgments}
This research is supported and implemented under the ARISTEIA action of the operational programme for education and long life learning and
is co-funded by the European Union (European Social Fund) and National Resources of Greece. A preliminary account of the results was
presented by one of us (G.P.) at the workshop ``Aspects of fluid/gravity correspondence" held in Thessaloniki, Greece, 16-20 February 2015.
We thank the organizers for their kind invitation and the participants for fruitful discussions.

\appendix

\section{Useful relations in \Poincare and Spherical Coordinates}
\label{sec:coordinate_transformation}

It is known that in \Poincare coordinates, the extremal surface corresponding to a disk region with radius $R$ centered at
$\left( x_0 , y_0 \right)$ is given by the ``hemisphere'',
\begin{equation}
\begin{split}
{\left( {x - {x_0}} \right)^2} + {\left( {y - {y_0}} \right)^2} + {z^2} &= {R^2},\\
\tau  &= {\tau _0} \, .
\end{split}
\label{eq:ct_surface_poincare}
\end{equation}
Changing to global spherical coordinates, using the coordinate transformation \eqref{eq:coordinate_transformation}, the surface \eqref{eq:ct_surface_poincare} takes the form,
\begin{equation}
\begin{split}
\cos \theta & = \cos {\theta _0}\sqrt {1 + \frac{1}{{{r^2}}}} \, ,\\
t &= {t_0} \, ,
\end{split}
\label{eq:ct_surface_global}
\end{equation}
where $\theta_0$ and $t_0$ are given by
\begin{align}
\cos {\theta _0} &= \frac{{1 - {R^2} + {\tau _0}^2}}{{\sqrt {\left[ {{{\left( {R + {\tau _0}} \right)}^2} + 1} \right]
\left[ {{{\left( {R - {\tau _0}} \right)}^2} + 1} \right]} }} \, ,\\
\cos {t_0} &= \frac{{1 + {R^2} - {\tau _0}^2}}{{\sqrt {\left[ {{{\left( {R + {\tau _0}} \right)}^2} + 1} \right]
\left[ {{{\left( {R - {\tau _0}} \right)}^2} + 1} \right]} }} \, .
\end{align}

To express the variation of the modular Hamiltonian for the polar cap region in spherical coordinates, we needed the
expression for the conformal Killing vector in the boundary, $\zeta$, that leaves the entangling curve invariant. This
is the limit of a conformal Killing vector in the bulk, $\xi$, which in \Poincare coordinates and with the appropriate
normalization takes the following form \cite{Faulkner:2013ica},
\begin{multline}
\xi  = \frac{\pi }{R}\left[ {\left( {{R^2} - {z^2} - {{\left( {\tau  - {\tau _0}} \right)}^2} - {{\left( {x - {x_0}} \right)}^2} - {{\left( {y - {y_0}} \right)}^2}} \right){\partial _\tau }} \right.\\
\left. \phantom{\left( {{{\left( {y - {y_0}} \right)}^2}} \right)} { - 2\left( {\tau  - {\tau _0}} \right)\left( {z{\partial _z} + \left( {x - {x_0}} \right){\partial _x} + \left( {y - {y_0}} \right){\partial _y}} \right)} \right] .
\end{multline}
In our analysis, the entangling curve is taken to be axially symmetric, as for the metric perturbations. This selects $x_0 = y_0 = 0$,
so that the center of the disc is placed at the north pole. Furthermore, since the $AdS_4$ metric is static, we may choose a specific
instant of time to simplify the calculations. The choice ${\tau _0} = \sqrt {1 + {R^2}}$ implies that
\begin{align}
\tan {\theta _0} &= R \, ,\\
\cos {t_0} &= 0 \, .
\end{align}
At that instant, the conformal Killing vector $\xi$ in the bulk takes the form
\begin{multline}
\xi  = \frac{\pi }{R}\left[ {\left( {{R^2} - {z^2} - {{\left( {\tau  - \sqrt {1 + {R^2}} } \right)}^2} - {x^2} - {y^2}} \right){\partial _\tau }} \right.\\
\left. \phantom{\left( {{{\left( {\tau  - \sqrt {1 + {R^2}} } \right)}^2}} \right)} { - 2\left( {\tau  - \sqrt {1 + {R^2}} } \right)\left( {z{\partial _z} + x{\partial _x} + y{\partial _y}} \right)} \right] .
\end{multline}

Converting to spherical coordinates by the transformation \eqref{eq:coordinate_transformation}, we find
\begin{multline}
\xi  = \frac{{2\pi \sqrt {1 + {R^2}} }}{R}\left[ {\left( {\frac{{r\cos \theta \sin t}}{{\sqrt {1 + {r^2}} }} - \frac{1}{{\sqrt {1 + {R^2}} }}} \right){\partial _t}} \phantom{\frac{{\sqrt {1 + {r^2}} }}{r}} \right.\\
\left. { - \sqrt {1 + {r^2}} \cos \theta \cos t{\partial _r} + \frac{{\sqrt {1 + {r^2}} \sin \theta \cos t}}{r}{\partial _\theta }} \right],
\end{multline}
which can be rewritten as
\begin{multline}
\xi  = \frac{{2\pi }}{{\sin {\theta _0}}}\left[ {\left( {\frac{{r\cos \theta \cos \left( {t - {t_0}} \right)}}{{\sqrt {1 + {r^2}} }} - \cos {\theta _0}} \right){\partial _t}} \phantom{\frac{{\sqrt {1 + {r^2}} \left( {t - {t_0}} \right)}}{r}} \right.\\
\left. { + \sqrt {1 + {r^2}} \cos \theta \sin \left( {t - {t_0}} \right){\partial _r} - \frac{{\sqrt {1 + {r^2}} \sin \theta \sin \left( {t - {t_0}} \right)}}{r}{\partial _\theta }} \right] .
\end{multline}

Then, the boundary limit of this expression provides the desired conformal Killing vector $\zeta$, which reads
\begin{equation}
\zeta = \frac{2 \pi}{{\sin {\theta _0}}}\left[ {\left( {\cos \left( {t - {t_0}} \right)\cos \theta  - \cos {\theta _0}} \right){\partial _t} - \sin \left( {t - {t_0}} \right)\sin \theta {\partial _\theta }} \right]
\end{equation}
and is used to compute the variation of the modular Hamiltonian in section \ref{subsec:dE_formula}.

\section{The Modular Flow for a Disc in Minkowski Space-time}
\label{sec:modular_Hamiltonian_disc}

It is not known how to express the modular Hamiltonian for a general region and state. This difficulty stems from the fact that,
in general, the modular Hamiltonian is a non-local operator. Exceptionally, in some cases, the modular Hamiltonian generates a geometric
flow and can be expressed as a local operator. Here, we outline how this is actually realized for a disk region in Minkowski space-time.

The modular Hamiltonian defines a symmetry of the system. More specifically, the symmetry group is provided by the unitary operators
$U\left( s \right) = e^{-iHs}$ and it is called the modular group,
\begin{equation}
\Tr\left( {\rho U\left( s \right)OU\left( { - s} \right)} \right) = \Tr\left( {\rho {\rho ^{is}}O{\rho ^{ - is}}} \right) = \Tr\left( {{\rho ^{ - is}}\rho {\rho ^{is}}O} \right) = \Tr\left( {\rho O} \right) .
\end{equation}
We define $O\left( s \right) = U\left( s \right) O U\left( -s \right)$. When the modular Hamiltonian is not a local operator,
this flow is non-local.

If one extends the modular group to imaginary parameters, there will be a periodicity relation in imaginary time,
\begin{equation}
\Tr\left( {\rho {O_1}\left( i \right){O_2}} \right) = \Tr\left( {\rho U\left( i \right){O_1}U\left( { - i} \right){O_2}} \right) = \Tr\left( {\rho {\rho ^{ - 1}}{O_1}\rho {O_2}} \right) = \Tr\left( {\rho {O_2}{O_1}} \right) .
\end{equation}
Thus, the state $\rho$ is thermal with respect to the time evolution of the modular symmetry with temperature $T=1$.

Let us consider Minkowski space-time in coordinates $X^\mu$. The Rindler space $\mathcal{R}$ is defined as the causal development
of the half plane $X^1 <0$. The causal development of a region $A$ is the set of space-time points whose causal space-time curves
necessarily cross region $A$. The modular transformations map the local operators $O\left( X \right)$ in Rindler space $\mathcal{R}$
to themselves. More specifically, the modular flow acts on the coordinates as follows,
\begin{equation}
{X^ \pm }\left( s \right) = {X^ \pm }{e^{ \pm 2\pi s}} , ~~~~~~ {X^i}\left( s \right) = {X^i} ,
\label{eq:modular_flow_old_coordinates}
\end{equation}
where ${X^ \pm } = {X^1} \pm {X^0}$ and the Latin indices $i$ run from 2 to $d-1$.

The modular Hamiltonian for $\mathcal{R}$ can be calculated explicitly. Transforming to the usual Rindler coordinates
${X^ \pm } = z{e^{ \pm \tau /R}}$, the metric becomes
\begin{equation}
d{s^2} =  - \frac{{{z^2}}}{{{R^2}}}d{\tau ^2} + d{z^2} + d{X^i}d{X^i} ,
\end{equation}
corresponding to a thermal state with temperature $T = 1 /2 \pi R$. Thus, the density matrix can be expressed as
\begin{equation}
{\rho _\mathcal{R}} = \frac{{{e^{ - 2\pi R{H_\tau }}}}}{{\Tr{e^{ - 2\pi R{H_\tau }}}}}
\end{equation}
with modular Hamiltonian
\begin{equation}
{H_\mathcal{R}} = 2\pi R{H_\tau } + \log \Tr{e^{ - 2\pi R{H_\tau }}} .
\end{equation}

The result for Rindler space can be subsequently used to obtain the modular Hamiltonian for a disc $D$ of radius $R$.
The conformal transformation
\begin{equation}
{x^\mu } = \frac{{{X^\mu } - {X_\nu }{X^\nu }{C^\mu }}}{{1 - 2{X_\kappa }{C^\kappa } +
{X_\rho }{X^\rho }{C_\sigma }{C^\sigma }}} + \frac{{{C^\mu }}}{{2{C_\lambda }{C^\lambda }}} \, ,
\end{equation}
with
\begin{equation}
{C^\mu } = \left( {0 \, , \, \frac{1}{{2R}} \, , \, 0 \, , \ldots , \, 0} \right)
\end{equation}
yields
\begin{equation}
{\eta _{\mu \nu }}d{X^\mu }d{X^\nu } = {\Omega ^2}{\eta _{\mu \nu }}d{x^\mu }d{x^\nu }
\end{equation}
with
\begin{equation}
\Omega  = 1 - 2{X_\kappa }{C^\kappa } + {X_\rho }{X^\rho }{C_\sigma }{C^\sigma } .
\end{equation}
Letting $r=\sqrt{\left( {x^1} \right)^2 + \ldots + \left( {x^{d-1}} \right)^2}$ and $t=x^0$, we see that this conformal
transformation maps the half plane $X^1 \le 0$ to the disk $D$, $r \le R$, and the Rindler space $X^\pm \le 0$ to the causal
development of the disk $\mathcal{D}$, $x^\pm \le R$, where $x^\pm = r \pm t$. For later use, we express the new radial coordinate as
\begin{equation}
r = \frac{1}{\Omega }\sqrt {{X^i}{X^i} + {{\left( {R - \frac{{{X_\nu }{X^\nu }}}{{4R}}} \right)}^2}} . \label{eq:conformal_transformation_r}
\end{equation}
Also, the inverse coordinate transformation is
\begin{equation}
{X^\mu } = \Omega \left( {{x^\mu } + 2{x_\nu }{x^\nu }{C^\mu }} \right) - \frac{{{C^\mu }}}{{{C_\lambda }{C^\lambda }}} \, ,
\label{eq:inverse_conformal_transformation}
\end{equation}
where
\begin{equation}
\Omega  = {\left( {\frac{1}{4} + {C_\kappa }{x^\kappa } + {C_\lambda }{C^\lambda }{x_\nu }{x^\nu }} \right)^{ - 1}} .
\end{equation}

We want to determine the form of the modular flow in the new coordinate system. For this purpose, we first make use of equations  \eqref{eq:modular_flow_old_coordinates} and \eqref{eq:inverse_conformal_transformation} to find the flow of the conformal factor $\Omega$,
\begin{equation}
\begin{split}
\Omega \left( s \right) &= 1 - \frac{{{X^1}\left( s \right)}}{R} + \frac{{{X_\nu }{X^\nu }}}{{4{R^2}}}\\
 &= \Omega  - \frac{1}{R}\left[ {\left( {\cosh 2\pi s - 1} \right){X^1} + \sinh 2\pi s{X^0}} \right]\\
 &= \Omega \left( {\frac{1}{2}\frac{{{R^2} + {r^2} - {t^2}}}{{{R^2}}} + \frac{1}{2}\cosh 2\pi s\frac{{{t^2} + {R^2} - {r^2}}}{{{R^2}}} - \sinh 2\pi s\frac{t}{R}} \right)\\
 &= \frac{\Omega }{{{R^2}}}\left( {R\cosh \pi s - {x^ + }\sinh \pi s} \right)\left( {R\cosh \pi s + {x^ - }\sinh \pi s} \right) .
\end{split}
\end{equation}
The radial coordinate $r$, which is given by equation \eqref{eq:conformal_transformation_r}, also flows through its dependence on the conformal
factor,
\begin{equation}
r\left( s \right) = \frac{1}{{\Omega \left( s \right)}}\sqrt {{X^i}{X^i} + {{\left( {R - \frac{{{X_\nu }{X^\nu }}}{{4R}}} \right)}^2}}
= \frac{\Omega }{{\Omega \left( s \right)}} \, r \, .
\end{equation}
Similarly, the modular flow of the new time coordinate can be expressed as
\begin{equation}
\begin{split}
t\left( s \right) &= \frac{{{X^0}\left( s \right)}}{{\Omega \left( s \right)}} = \frac{{\cosh 2\pi s{X^0} +
\sinh 2\pi s{X^1}}}{{\Omega \left( s \right)}}\\
 &= \frac{\Omega }{{\Omega \left( s \right)}} \, R\left( {\cosh 2\pi s\frac{t}{R} - \frac{1}{2}\sinh 2\pi s\frac{{{t^2} +
 {R^2} - {r^2}}}{{{R^2}}}} \right) ,
\end{split}
\end{equation}
which in turn implies that
\begin{equation}
\begin{split}
\frac{{\Omega \left( s \right)}}{\Omega } \, {x^ \pm }\left( s \right) &= r \pm R\left( {\cosh 2\pi s\frac{t}{R} -
\frac{1}{2}\sinh 2\pi s\frac{{{t^2} + {R^2} - {r^2}}}{{{R^2}}}} \right)\\
 &= \frac{1}{R}\left( {R\cosh \pi s \pm {x^ \mp }\sinh \pi s} \right)\left( {{x^ \pm }\cosh \pi s \mp R\sinh \pi s} \right) .
\end{split}
\end{equation}
Finally, we obtain
\begin{equation}
{x^ \pm }\left( s \right) = R\frac{{{x^ \pm }\cosh \pi s \mp R\sinh \pi s}}{{R\cosh \pi s \mp {x^ \pm }\sinh \pi s}} =
R\frac{{\left( {R + {x^ \pm }} \right) - \left( {R - {x^ \pm }} \right){e^{ \mp 2\pi s}}}}{{\left( {R + {x^ \pm }} \right) +
\left( {R - {x^ \pm }} \right){e^{ \mp 2\pi s}}}} .
\end{equation}

The modular flow in the new coordinates yields by differentiation
\begin{align}
{\left. {\frac{{\partial {x^ \pm }\left( s \right)}}{{\partial s}}} \right|_{s = 0}} &=  \pm \pi \frac{{{R^2} - {x^ \pm }^2}}{R} \, ,\\
{\left. {\frac{{\partial r\left( s \right)}}{{\partial s}}} \right|_{s = 0}} &= - 2\pi \frac{{rt}}{R} \, ,\\
{\left. {\frac{{\partial t\left( s \right)}}{{\partial s}}} \right|_{s = 0}} &= \pi \frac{{{R^2} - {r^2} - {t^2}}}{R} \, ,
\end{align}
clarifying the action of the modular Hamiltonian on local operators in the new coordinates. Restricting attention to
the time slice $t = 0$, one recovers the usual expression for the modular Hamiltonian for a disk of radius $R$ \cite{Casini:2011kv}, namely
\begin{equation}
H_D = 2\pi \int {{d^{d - 1}}x \, \frac{{{R^2} - {r^2}}}{{2R}} \, {\mathcal{T}^{00}}} .
\end{equation}

\section{Coordinate Dependence of Subleading Terms in $\delta A_g^{\rm extr}$, $\delta A_b^{\rm extr}$}
\label{sec:coordinate_dependence}

The individual contributions to the variation of the area, $\delta A_g^{\rm extr}$ and $\delta A_b^{\rm extr}$, which appear in
section \ref{sec:EE_perturbed_AdS4_calculation}, are coordinate dependent,
unlike their sum which is coordinate independent when the line element of the entangling curve remains constant. Furthermore, the term
$\delta A_b^{\rm extr}$ only vanishes for Dirichlet boundary conditions provided that one works in coordinate systems, unlike ours,
where specific metric elements (to be made more precise later) contain no contributions that fall like $1/r$ close to the boundary.
These statements are substantiated below by performing a small perturbative coordinate transformation altering the $1/r$ asymptotics of the
metric elements,
\begin{equation}
r = \sqrt {r{'^2} + \epsilon \left( t , \theta \right) r'} .
\label{eq:asymptotic_CT}
\end{equation}

The area of the extremal surface in $AdS_4$ is provided by equation \eqref{eq:unperturbed_area}. Writing the result
in terms of the $r' $ coordinate, we have
\begin{equation}
\begin{split}
A^{\rm extr} &= 2\pi \mathop {\lim }\limits_{r' \to \infty } \left( {\sqrt {r{'^2} + \varepsilon
\left( {{t_0},\theta \left( r \right)} \right)r'} \, \sin {\theta _0} - 1} \right) \\
&= 2\pi \mathop {\lim }\limits_{r' \to \infty } \left( {\left( {r' + \frac{{\varepsilon \left( {{t_0},{\theta _0}} \right)}}{2}} \right)
\sin {\theta _0} - 1} \right) .
\end{split}
\end{equation}
The change of coordinate introduces a change in the area of the minimal surface in $AdS_4$, which functionally resembles $\delta A_g^{\rm extr}$,
\begin{equation}
\delta A{'_g}^{\rm extr} = \pi \varepsilon \left( {{t_0},{\theta _0}} \right)\sin {\theta _0} .
\end{equation}
This seems to be odd at first sight, because the area should be independent of the choice of coordinates, but $A^{\rm extr}$ is infinite and the result depends on way one is taking the limit. Ultimately we are interested in finding a result that is independent of the limiting process.

On the other hand, the metric element ${g_{\varphi \varphi }}$ of $AdS_4$ space in the $r'$ coordinate is expressed as
\begin{equation}
{g_{\varphi \varphi }} = \left( {r{'^2} + \varepsilon \left( {{t},{\theta}} \right)r'} \right){\sin ^2}{\theta} \, .
\end{equation}
Demanding that the length of the entangling curve remains invariant, we obtain
\begin{equation}
\delta \theta \left( {{t_0}} \right) =  - \frac{1}{{2r'}}\varepsilon \left( {{t_0},{\theta _0}} \right)\tan {\theta _0} \, ,
\end{equation}
which, in turn, gives a non-vanishing contribution to the area of the minimal surface, which resembles $\delta {A_b}^{\rm extr}$,
\begin{equation}
\delta A{'_b}^{\rm extr} =  - \pi \varepsilon \left( {{t_0},{\theta _0}} \right)\sin {\theta _0} \, .
\end{equation}

Thus, demanding that the line element of the entangling curve remains constant, we arrive at the following result,
\begin{equation}
\delta A{'_g}^{\rm extr} + \delta A{'_b}^{\rm extr} = 0 .
\end{equation}
This provides the desired coordinate independent prescription for taking the limit. The variation of the area of the minimal surface in $AdS_4$ should be zero under changes of coordinates. Likewise, for perturbations of $AdS_4$ space-time, the combined variation $\delta A_g^{\rm extr} + \delta A_b^{\rm extr}$, which is not zero, does not depend on the choice of coordinates, unlike the individual terms $\delta A_g^{\rm extr}$ and $\delta A_b^{\rm extr}$ that depend upon it.

As an illustrative example of the effect that coordinate transformations may have to the components of the variation of the area of the minimal surface in the presence of gravitational perturbations of $AdS_4$ space-time, we consider the case of polar perturbations. Using the change of coordinate \eqref{eq:asymptotic_CT} with the choice
\begin{equation}
\varepsilon \left( {t,\theta } \right) = {e^{ - i\omega t}}{J_0}\left( {\frac{1}{2}l\left( {l + 1} \right) - {\omega ^2}} \right)
{P_l}\left( {\cos \theta } \right),
\end{equation}
we find
\begin{multline}
\delta {A'_g}^{\rm extr} =  - \pi {e^{ - i\omega {t_0}}}\mathop {\lim }\limits_{r' \to \infty }
\left[ {\left( {r'{J_1} + \frac{1}{2}{J_0} \, l\left( {l + 1} \right)} \right)\sin {\theta _0} \,
{P_l}\left( {\cos {\theta _0}} \right)} \right.\\
\left. { + \cot {\theta _0} \, \frac{{l\left( {l + 1} \right)}}{{2l + 1}}{J_0}\left( {{P_{l + 1}}
\left( {\cos {\theta _0}} \right) - {P_{l - 1}}\left( {\cos {\theta _0}} \right)} \right)} \right]
\end{multline}
and
\begin{equation}
\delta {A'_b}^{\rm extr} = \pi {e^{ - i\omega {t_0}}}\mathop {\lim }\limits_{r' \to \infty } r'{J_1}\sin {\theta _0} \,
{P_l}\left( {\cos {\theta _0}} \right).
\end{equation}
One can readily verify that $\delta {A'_g}^{\rm extr} + \delta {A'_b}^{\rm extr} = \delta {A_g}^{\rm extr} + \delta {A_b}^{\rm extr}$, using the results of section \ref{sec:EE_perturbed_AdS4_calculation}, as advertised.

Choosing Dirichlet boundary conditions for the metric, $J_1 = 0$, we find that $\delta {A'_b}^{\rm extr}$ vanishes, whereas
$\delta {A'_g}^{\rm extr}$ is finite, equal to $4 \pi \delta E$. Note at this end that the asymptotic behavior of ${g_{\varphi \varphi }}$ in $r'$ coordinate is
\begin{equation}
{g_{\varphi \varphi }} = r{'^2}{\sin ^2}{\theta _0}\left( {1 + \mathcal{O}\left( {\frac{1}{{{r'^2}}}} \right)} \right).
\end{equation}

Summarizing, the term $\delta A_b^{\rm extr}$ can only be neglected for Dirichlet boundary conditions and in
coordinate systems where the metric elements corresponding to elementary lengths parallel to the entangling surface do not contain
subleading terms that fall like $1/r$ close to the boundary. This is precisely the content of the calculations reported in earlier works \cite{Lashkari:2013koa,Faulkner:2013ica}, using Dirichlet boundary conditions and writing the metric in Fefferman-Graham
coordinates which obey the restrictions stated above. If any one of these restrictions do not hold, the variance $\delta A_b^{\rm extr}$
will become relevant in the calculation and the need for invariance of the line element of the entangling surface will come into play,
as in our calculations. As emphasized above, the individual terms $\delta {A_g}^{\rm extr}$ and $\delta {A_b}^{\rm extr}$ contain gauge artifacts, but not their sum, which is coordinate independent.

%\section*{References}

\end{document}